\documentclass[a4paper,11pt]{article}
\pdfoutput=1 

\usepackage{jcappub} 

\usepackage{subcaption}
\usepackage{mwe}
\usepackage[T1]{fontenc} 
\usepackage{xcolor}

\usepackage{xspace}
\usepackage{siunitx}
\newcommand{\kevee}{keV_{ee}}
\newcommand{\Cygnus}{\textsc{Cygnus}\xspace}

\title{Observables for Recoil Identification in High-Definition Gas Time Projection Chambers}

\author[a,1]{M. Ghrear,\note{Corresponding author.}}
\author[a]{S.E. Vahsen,}
\author[b]{C. Deaconu}

\affiliation[a]{Department of Physics and Astronomy, University of Hawaii,\\2505 Correa Road, Honolulu, HI, 96822, USA}
\affiliation[b]{Department of Physics, Enrico Fermi Inst., Kavli Inst. for Cosmological Physics, University of Chicago,\\ 5640 S Ellis Ave, Chicago, IL 60637, USA}

\emailAdd{majd@hawaii.edu}
\emailAdd{sevahsen@hawaii.edu}

\abstract{Directional detection of nuclear recoils is broadly desirable in nuclear and particle physics. At low recoil energies, this capability may be used to confirm the cosmological origin of a dark matter signal, to penetrate the so-called neutrino floor, or to distinguish between different neutrino sources. Gas Time Projection Chambers (TPCs) can enable directional recoil detection if the readout granularity is sufficiently high, as is the case when micro-pattern gaseous detectors (MPGDs) are utilized. A key challenge in such detectors is identifying and rejecting background electron recoil events caused by gamma rays from radioactive contaminants in the detector materials and the environment. We define new observables that can distinguish electron and nuclear recoils, even at keV-scale energies, based on the simulated ionization’s topology. We perform a simulation study that shows these observables outperform the traditionally used discriminant, dE/dx, by up to three orders of magnitude. Furthermore, these new observables work well even at ionization energies well below \SI{10}{keV} and remain robust even in the regime where directionality fails.}

\begin{document}
\maketitle
\flushbottom

\section{Introduction}
\label{sec:introduction}
Directional dark matter detection was first proposed in 1988 by D. N. Spergel~\cite{spergel1988motion}. In the Standard Dark Halo Model, the Earth is expected to experience an approximately \SI{220}{km/s} wind of dark matter due to the Milky Way's rotation through the dark matter halo~\cite{DMmodel}. This mechanism gives Galactic dark matter a unique directional signature that can distinguish a WIMP-induced signal from backgrounds. Constructing a direct dark matter detection experiment that has both relevant target mass and suitable directional performance is highly challenging yet ultimately appears feasible. Reviews of past and ongoing efforts can be found in refs.~\cite{a,b,c,sven_review}. At this time, efforts based on gas time projection chambers (gas TPCs)~\cite{d,e,f,g,h,h2,i} and nuclear emulsions~\cite{j} appear most advanced. All directional experiments that have set dark matter limits to date utilize gas TPCs. \Cygnus~\cite{cygnus} is a new effort where most of these gas TPC groups intend to construct a world-wide network of detectors for directional dark matter searches with sufficient target mass. Gas TPCs capable of directional dark matter searches are widely applicable, of interest for example also in neutron detection, neutrino physics, and for precision measurements of nuclear recoils themselves~\cite{sven_review}.

One of the main obstacles faced by the use of gas TPCs for dark matter searches is the discrimination of background electron recoils~\cite{lopez2012background,battat2014radon,billard2012low,lopez2012rejection,dinesh}.  In a recent  \Cygnus analysis~\cite{cygnus}, a flat predicted electron background spectrum combined with exponential electron rejection indicated that the energy threshold of a future gas TPC experiment (if requiring background-free operation) would be determined by electron rejection. Gas TPCs with high readout segmentation provide three-dimensional recoil track information that can be used to distinguish electron recoils from nuclear recoils. As a particle recoils through a gas, it undergoes multiple scattering with the constituents of the gas. As a result, the particle's path becomes altered in a way that depends on the pressure and constituents of the gas, as well as the energy and the type of the particle. Hence, it is expected that the reconstructed tracks for nuclear and electron recoils have characteristic differences.

Previous work on electron rejection can be found in refs.~\cite{dinesh,billard2012low,MIMAC2,CYGNO_erej}. While dE/dx has been the standard way to identify electrons, several of these studies develop their own observables and methodologies. It is not straightforward to conduct a one-to-one comparison on electron rejection results. In Section~\ref{Discussion}, we discuss important considerations and compare our findings to previous work. Our approach emphasizes the importance of studying electron rejection versus energy and all of our electron rejection results are given at specific energies.
 
In this paper, we investigate seven observables that identify characteristic differences between nuclear and electron recoils to distinguish them. In Section~\ref{sec:method}, we detail the methods used to simulate nuclear and electron recoils and the method for assessing our observables. In Section~\ref{sec:define}, we define the observables. In defining them, we demonstrate how their definition may be optimized for energy scales where nuclear recoils exhibit strong and weak directionality. Furthermore, in Section~\ref{sec:results} we assess the electron rejection achieved by our observables in the directional and weakly-directional regimes. Finally, in Section~\ref{sec:conclusion} we summarize our results. We also suggest future studies to improve electron rejection with these observables further. Our study is part of wider gas TPC detector R\&D efforts at the University of Hawaii, which include both the BEAST neutron detection project~\cite{Jaegle}, the initial Directional Dark Matter Detector project ($\rm D^3$ project)~\cite{9,10,11,h,13}, and more recently \Cygnus~\cite{cygnus}.

\section{Method}
\label{sec:method}

Our main goal is to define new observables that improve discrimination between electron and nuclear recoils. We aim for robust observables that characterize physical aspects of the ionization distribution. Such observables should be widely applicable, though perhaps with application-specific input parameters and selection values, across different gases and detector technologies. To illustrate this aspect, how the same observables can be re-optimized for different applications, we show results for a single gas mixture, but for two nuclides and energies, corresponding to strong and weak directionality.

\subsection{Choice of gas and detector parameters}
\label{choice}

Even though we do not advocate for a specific gas, we must pick a specific mixture and electrical drift-field strength. These choices then determine the drift velocity and the diffusion of drift charge in the detector, both of which strongly affect particle identification capabilities. We choose to simulate a $80 \% \textrm{ He} + 10 \% \textrm{ CF}_4 + 10 \% \textrm{ CHF}_3$ mixture at a total pressure of \SI{60}{Torr} at \SI{25}{\degree C}. This electron drift gas mixture, where ionization is transported via free electrons, was specifically selected to have a similar value of $1 / ( \rho \cdot \sigma_T )$ (where $\rho$ is density and $\sigma_T$ is transverse diffusion) as the atmospheric-pressure (740 torr : 20 torr ) He:SF$_6$ negative ion drift gas (NID) studied in the \Cygnus experiment proposal~\cite{cygnus}. In a negative ion gas, the primary ionization attaches to and is transported by negative ions rather than electrons. Since recoil-length (L) is proportional to $1/\rho$, $1 / ( \rho \cdot \sigma_T )$ is proportional to the ratio of recoil-length to diffusion, which largely determines the angular resolution for nuclear recoils and strongly impacts the electron rejection capabilities. The quantity $1 / ( \rho \cdot \sigma_T )$ can thus be considered a figure of merit for design optimization, with higher values generally improving detector performance. Specific values for our study and a few others can be found in Table~\ref{tab:2}.

In ref.~\cite{cygnus}, it was found that lowering the density further, for example by going to (755:5) He:SF$_6$, and improving electron rejection by utilizing more advanced observables, are most likely both required. Here we focus on the second option. Going to lower gas density should lead to even further improvements, but we here choose a gas density that we think is experimentally achievable with a low-pressure electron gas mixture. Like the \Cygnus gas, our gas also contains both helium and fluorine nuclei, with the former improving spin-independent reach to low-mass WIMPs, and the latter improving sensitivity to spin-dependent WIMP-nucleon scattering. There are three short-term advantages to choosing an electron drift gas: Gas avalanche gains are higher, calculations of gas parameters via standard tools are generally reliable at the percent level, and existing readout electronics have been optimized for the drift velocities in such gases. Together, this means that the next stage of our study, experimental demonstration of improved electron rejection, will be simplified and can be performed sooner with an electron drift gas. The biggest drawback with utilizing electron drift, is the increased diffusion compare to NID. It is due to the increased diffusion that we must utilize sub-atmospheric pressure with electron drift gases, increasing the recoil length $L$, to achieve satisfactory $1 / ( \rho \cdot \sigma_T )$. Due to the reduced diffusion, NID may enable acceptable angular resolution for nuclear recoils even with atmospheric-pressure operation, which would lead to improved directional dark matter reach at reduced cost. But this development is expected to occur on a slightly longer timescale.

Because our study is meant to be agnostic with respect to TPC readout technology, we only simulate the diffusion due to drift, and then bin ionization after drift assuming a \SI{100}{\micro m} readout segmentation in each of the two readout plane dimensions ($x,y$) and in the drift direction ($z$). We note that our binning is slightly larger than the current state-of-the-art charge readout technology~\cite{GRIDPIX_NID}, meaning that it is already experimentally achievable. We simulate a third gas component, $\textrm{ CHF}_3$, and a relatively low drift field value, $40.6$ V/cm, in order to fine-tune the drift velocity, so that for typical readout electronics operating at \SI{40}{MHz} (25 ns time binning), the quantization of \SI{100}{\micro m} in the drift direction is obtained. The $\textrm{ CHF}_3$ component has been used by the MIMAC collaboration, which has already experimentally demonstrated the use of this gas additive, and found good agreement between measured and simulated drift velocities~\cite{MIMAC_DRIFT_VELOCITY}.

\subsection{Simulating Recoils}
\label{sims}

For the primary recoiling nuclei, we restrict our attention to helium, due to its long directional recoil tracks, and fluorine, because of its spin-dependent sensitivity. To simulate the recoiling nuclei we use \texttt{SRIM}~\cite{srim} and we post-process the  \texttt{SRIM} results using \texttt{retrim}~\cite{retrim} in order to obtain the charge distributions. The average energy per electron-ion pair ($W$ value) and the Fano factor ($\mathcal{F}$) required in \texttt{retrim} were calculated using \texttt{Garfield++/Heed}~\cite{heed} as $W=35.0$ eV and $\mathcal{F}=0.19$, respectively. The \texttt{SRIM} calculation requires a compound correction for each possible recoiling nucleus (primary and secondary). A compound correction is a simple way to approximate changes to the use of Bragg's rule, based on the compound being used. The compound dictionary in \texttt{SRIM} provides corrections for individual compounds, but it does not include any averaging for mixtures of compounds. To obtain an appropriate correction for our mixture, we average the individual compound corrections with respect to the probability of interaction. Such an approach has been used in ref.~\cite{avg} to compute an average W value for a gas mixture; we adopt the same approach, except using it to obtain an average compound correction for our mixture as 
\begin{equation}
C_i = \frac{f_1 \left[ \frac{dE}{dx} \right]_1 + f_2 \left[ \frac{dE}{dx} \right]_2 + f_3 \left[ \frac{dE}{dx} \right]_3 }{ \frac{f_1}{C_{1i}} \left[ \frac{dE}{dx} \right]_1 +  \frac{f_2}{C_{2i}}  \left[ \frac{dE}{dx} \right]_2 +  \frac{f_3}{C_{3i}}  \left[ \frac{dE}{dx} \right]_3 }.
\label{correct}
\end{equation}
In the above equation, the index $i = \{ \textrm{He}, \textrm{H},  \textrm{ F}, \textrm{C} \} $ specifies the recoiling nucleus. Furthermore, $C_i$ is the averaged compound correction and $C_{1i}$, $C_{2i}$, $C_{3i}$ are the compound corrections for the specified nucleus in He, $\textrm{ CF}_4 $, and $ \textrm{ CHF}_3$ respectively. The values $f_1$, $f_2$, $f_3$ are the molecular fractions of the gases. The energy loss per unit length $\left[ dE/dx \right]_1$  and $\left[ dE/dx \right]_2$ are obtained from ref.~\cite{de/dx} (and have been rescaled to a total pressure of  \SI{60}{Torr} and temperature of \SI{25}{\degree C}) for He and  $\textrm{ CF}_4$. Since the energy loss per unit length is approximately proportional to electron density, for $ \textrm{ CHF}_3$ we approximate $\left[ dE/dx \right]_3 \approx \left[ dE/dx \right]_2 \times \frac{40}{48}$. All of the values relevant to Eq.~\ref{correct} are summarized in Table~\ref{tab:1}.

\begin{table}[tbp]
\centering
\begin{tabular}{|p{10mm}||p{15mm}|p{15mm}|p{15mm}|p{15mm}|p{15mm}|p{15mm}|p{15mm}|}
\hline
Gas & Fraction & [dE/dx] (keV/cm) & Comp. Corr. He  & Comp. Corr. H  & Comp. Corr. C & Comp. Corr. F  \\
\hline
$\textrm{He}$ & 0.80 & 0.026 & 1.00 & 1.00 & 1.00 & 1.00 \\
$\textrm{CF}_4$ & 0.10 & 0.56 & 0.957 & 0.959 & 0.974 & 0.974 \\
$\textrm{CHF}_3$ & 0.10  & 0.47 & 0.969 & 0.965 & 0.991 & 0.991 \\
\hline
\end{tabular}
\caption{\label{tab:1} Parameters used to simulate our gas mixture. The $\left[ dE/dx \right]$ values for He and $\textrm{ CF}_4$ were obtained from ref.~\cite{de/dx} (after rescaling to a total pressure of  \SI{60}{Torr} and temperature of \SI{25}{\degree C}) and it was calculated for  $ \textrm{ CHF}_3$ as explained in the text. The compound corrections were obtained from \texttt{SRIM}~\cite{srim}.}
\end{table}

\texttt{DEGRAD}~\cite{degrad} offers the most complete simulation of electron recoils in gases~\cite{degradbest}; however, since it is not capable of simulating recoils in gas mixtures containing $\textrm{CHF}_3$, some special treatment is required. Based on advice by the DEGRAD developer S. Biagi, we reason that since the energy loss per unit length is approximately proportional to electron density, recoil tracks in $\textrm{CHF}_3$ will be slightly longer than those in $\textrm{ CF}_4$. Hence, replacing the number density of $\textrm{CHF}_3$ in our mixture with $\frac{40}{48}$ times that number density and using $\textrm{ CF}_4$ instead will result in charge distributions with approximately the same length. This method may be of general interest since the MIMAC collaboration also utilizes $\textrm{CHF}_3$ gas~\cite{mimacgas}, and it is crucial to be able to simulate the electron background accurately for low energy searches. 

\begin{figure}[tbp]
\centering 
\includegraphics[width=.45\textwidth]{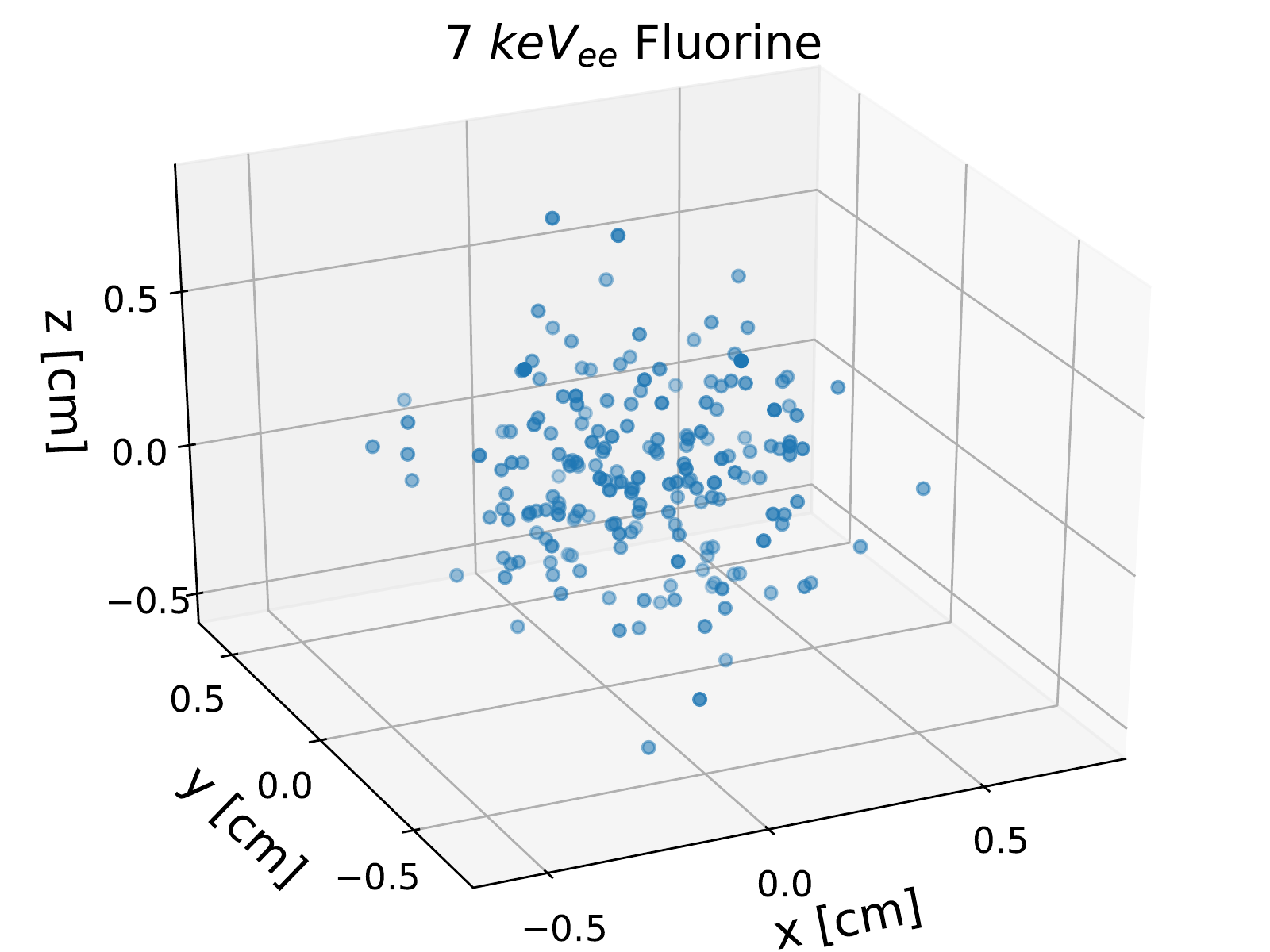}
\hfill
\includegraphics[width=.45\textwidth]{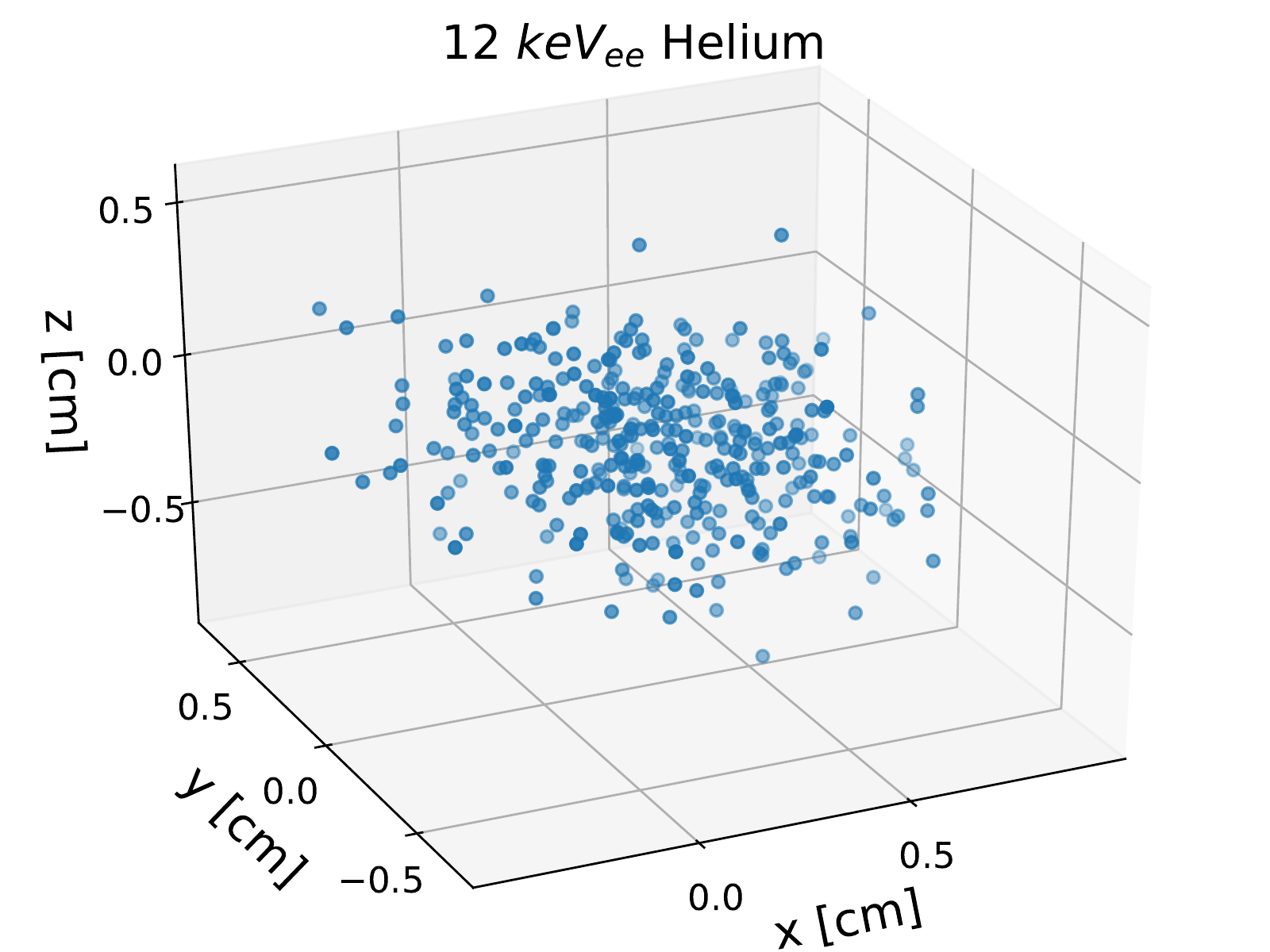}
\hfill
\includegraphics[width=.45\textwidth]{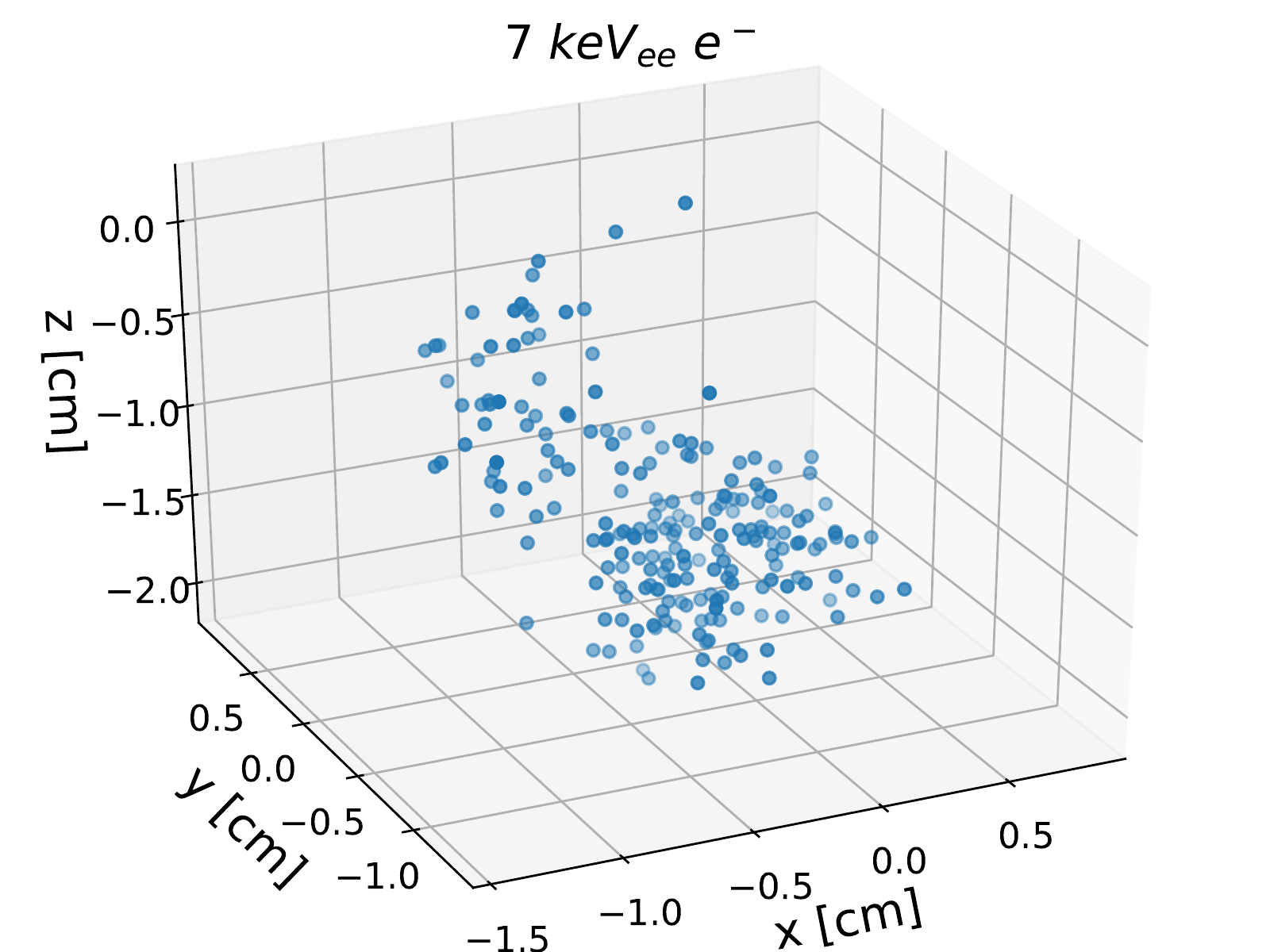}
\hfill
\includegraphics[width=.45\textwidth]{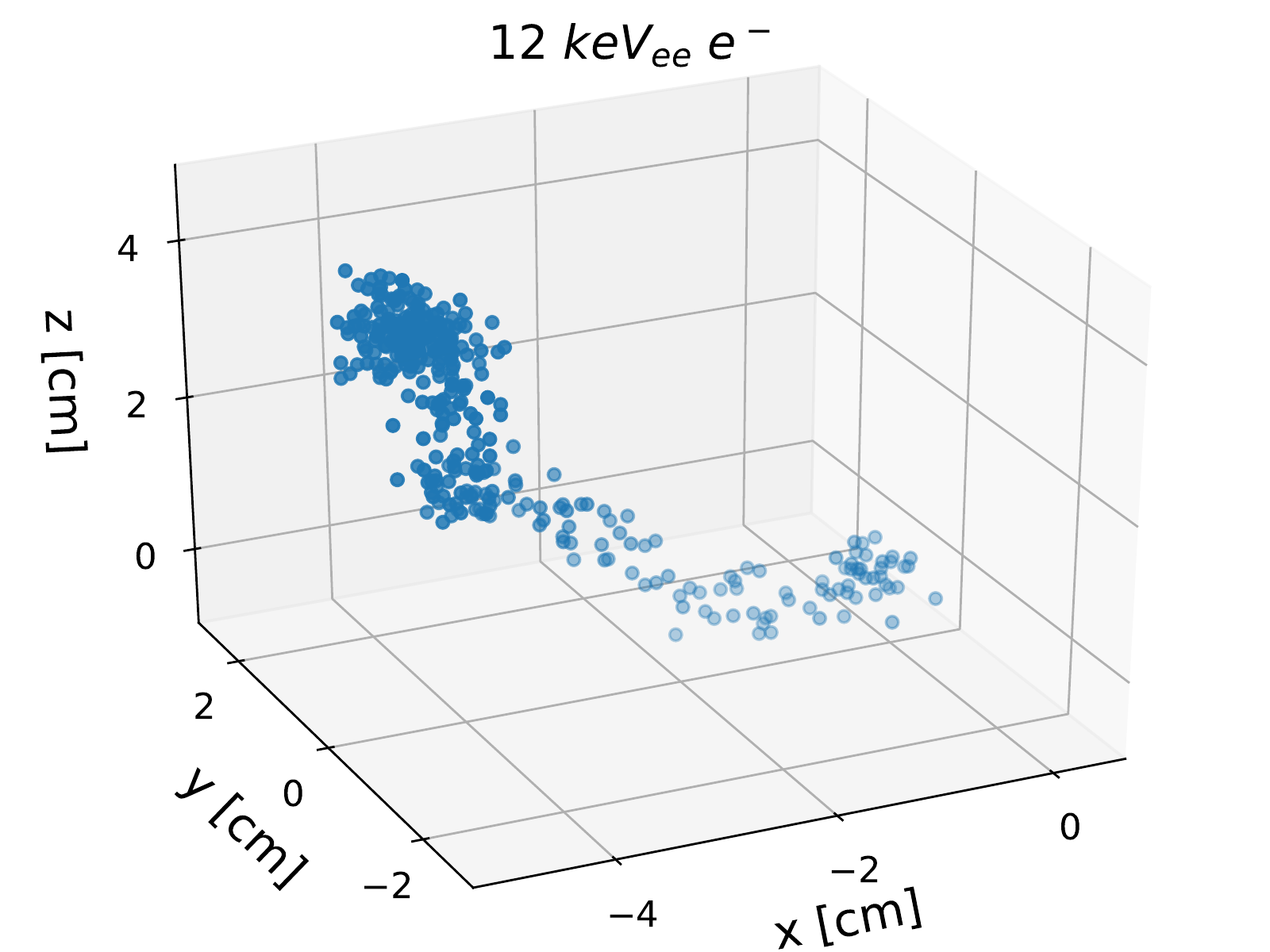}
\caption{\label{fig:tracks} Recoil tracks generated using \texttt{DEGRAD}, \texttt{SRIM}, and \texttt{retrim} as described in Section~\ref{sims}. Simulation assumes \SI{25}{cm} of drift in a $80 \% \textrm{ He} + 10 \% \textrm{ CF}_4 + 10 \% \textrm{ CHF}_3$ gas mixture at total pressure of $60$ Torr,  a temperature of \SI{25}{\degree C}  and in a $40.6$ V/cm drift field. The top and bottom left are fluorine and $e^-$ recoils in the \SI{7}{\kevee} energy bins, respectively. The top and bottom right are helium and $e^-$ recoils in the \SI{12}{\kevee} energy bins, respectively. }
\end{figure}

The \texttt{DEGRAD} simulations are isotropic; however, the \texttt{SRIM} and \texttt{retrim} simulations require an additional step to randomize their recoil directions. After obtaining an isotropic set of simulations, we apply the longitudinal and transverse diffusion assuming a drift length of \SI{25}{cm} (which results in diffusion greater than the average diffusion expected in a detector with 50 cm drift length, as proposed for the \Cygnus experiment~\cite{cygnus}). Using \texttt{Magboltz}~\cite{magboltz}, the transverse and longitudinal diffusion coefficients for our setup are calculated as $\sigma_T = 398$ $\si\micro$m/$\sqrt{\textrm cm}$ and  $\sigma_L = 425$ $\si\micro$m/$\sqrt{\textrm cm}$, respectively. The final step is to bin the recoil tracks into  \SI{100}{\micro m} $\times$ \SI{100}{\micro m} $\times$ \SI{100}{\micro m}  $\si\micro$m pixels. In our binning, we assume that every electron is counted. Several recoils generated in this manner are shown in Figure~\ref{fig:tracks}.

\subsection{Assessing Electron Rejection}
\label{assessER}

We need a methodology to measure how well an observable distinguishes nuclear recoils from electron recoils with the same total detected charge. The electron-equivalent energy ($E_{ee}$) of a recoil is defined as the recoil energy of an electron, which would produce the same total detected charge as said recoil. We work in terms of electron-equivalent units because it allows us to intuitively assess how well our observables reject electrons of a given energy. We need the average energy per electron-ion pair created by an electron recoil ($W_e$) to determine the electron-equivalent energy. Using \texttt{Degrad} we determine ($W_e$) by simulating electron recoils at the energies we are interested in (3-\SI{15}{keV}). Then for each track, we divide the recoil energy by the total number of ionized electrons. Using this method, we find that $W_e$ is nearly energy independent for the 3-\SI{15}{keV} energy range, taking on values within $W_e = \SI{32.4 \pm 0.1}{eV}$. Moving forward, we simply use $W_e = \SI{32.4}{eV}$. With this quantity, we may calculate the electron-equivalent energy of a specific recoil as
\begin{equation}
E_{ee} = N_e \times W_e,
\label{Ienergy}
\end{equation}
where $N_e$ is the number of ionized electrons in the recoil. 

\begin{figure}[tbp]
\centering 
\includegraphics[width=.8\textwidth]{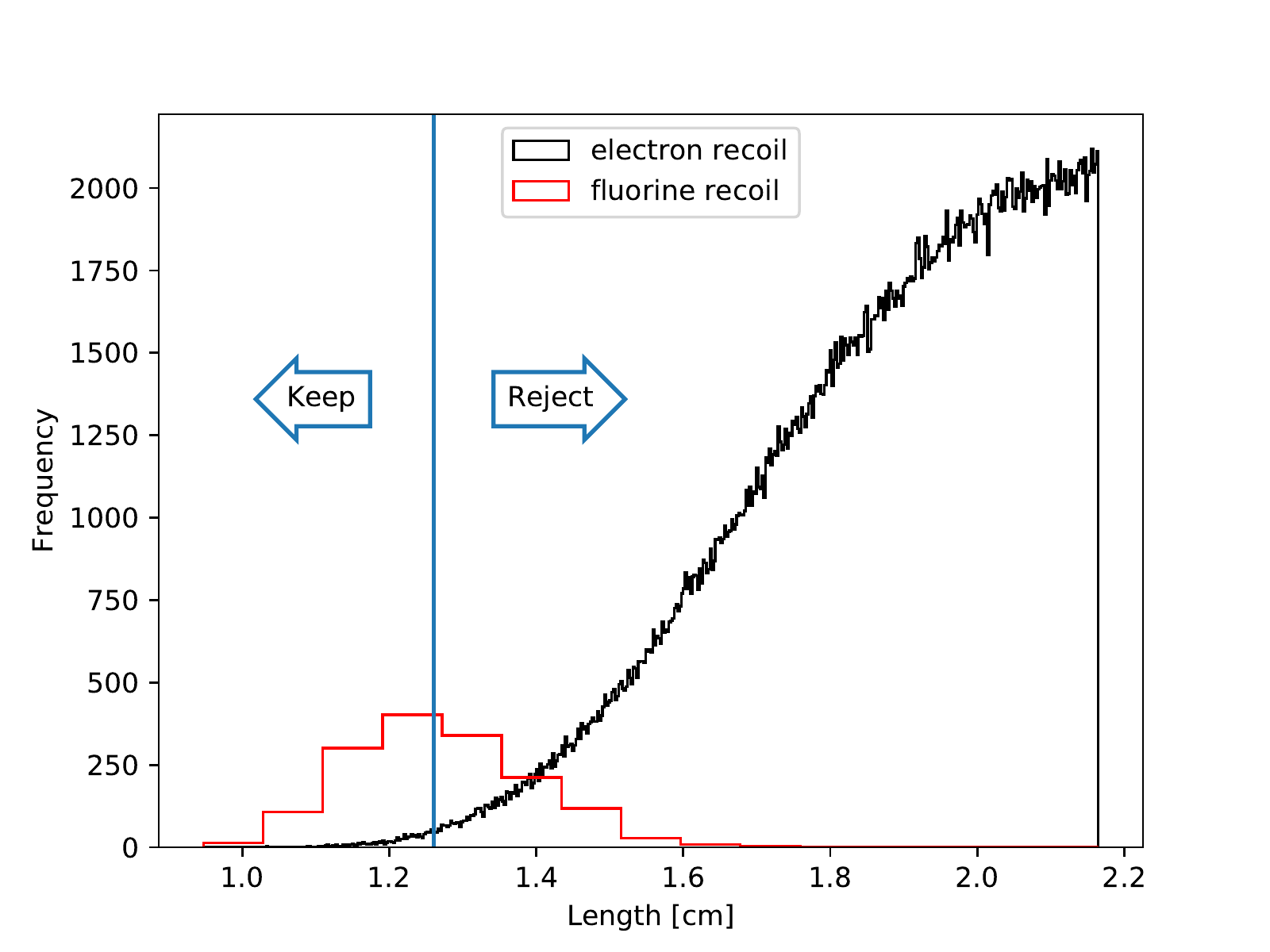}
\hfill
\caption{\label{fig:distribs} Length Along Principal Axis for  \SI{7}{\kevee}  electron recoils (black) and fluorine recoils (red). Recoils simulated after \SI{25}{cm} of drift in a $80 \% \textrm{ He} + 10 \% \textrm{ CF}_4 + 10 \% \textrm{ CHF}_3$ gas mixture at total pressure of $60$ Torr,  a temperature of \SI{25}{\degree C}  and in a $40.6$ V/cm drift field. The blue vertical line corresponds to a nuclear recoil selection efficiency of 0.5. All recoils beyond the vertical line are rejected.}
\end{figure}

We bin the electron-equivalent energies into \SI{1}{\kevee} bins. For each bin, the performance of an observable is determined by its electron rejection factor, defined as
\begin{equation}
R \equiv \frac{N^{total}_e}{N^{kept}_e}.
\label{rejection}
\end{equation}
Above, $N^{total}_e$ is the total number of electron recoils analyzed, and $N^{kept}_e$ is the number of those recoils that the observable failed to reject.  The criteria we use to reject electrons is determined by the desired nuclear recoil selection efficiency ($\epsilon$), defined as
\begin{equation}
\epsilon  \equiv \frac{N_{He}^{kept}}{N_{He}^{total}},
\label{NucEff}
\end{equation}
where $N_{He}^{total}$ is the total number of helium recoils and $N_{He}^{kept}$ is the number of nuclear recoils that are kept. Each observable yields a distribution of values for the electron and nuclear recoils within the energy bin, and $\epsilon$ specifies a cutoff at which we reject all recoils. An example of this process is illustrated in Figure~\ref{fig:distribs} for fluorine recoils in the \SI{7}{\kevee}  energy bin.
In summary, for a set of simulated electron and nuclear recoils in an electron-equivalent energy bin:
\begin{enumerate}
\item We choose a desired nuclear recoil selection efficiency.
\item Based on the distribution of an observable's values for the nuclear recoils, a threshold is obtained.
\item Electrons are rejected based on the cutoff from step 2, and the electron rejection factor is calculated.
\item Step 1-3 are repeated for several nuclear recoil selection efficiency values.
\end{enumerate}
All electron rejection plots utilize asymmetric  Poisson error bars appropriate for efficiencies. These errors include statistical uncertainties resulting directly from finite electron recoil statistics, and indirectly from finite nuclear recoil samples, which lead to an uncertainty in the discriminant selection values required to achieve the desired nuclear recoil selection efficiency. Both effects are added in quadrature.

\section{Observables}
\label{sec:define}

We will define several observables that distinguish electron recoils from nuclear recoils using the detailed charge cloud topology information accessible to gaseous TPCs with highly-segmented readouts. Several of our observables are optimizable for the specific gas mixture used and energy scales of interest. In directional gas TPCs, the angular resolution for nuclear recoils depends strongly on recoil energy. Typically, at the detection threshold, recoils have no directionality and the directionality gradually turns on as energy increases. For the purpose of this study, we only optimize our observables for two scenarios. The first scenario is where the recoil track length is long with respect to diffusion. In this case, we are able to accurately infer the direction of the recoil, and so we call this the directional optimization. The second scenario happens at lower energies (but still well above detection threshold), where the recoil track length is of comparable scale to the diffusion. Here, it is no longer possible to accurately determine the direction of the recoil and so we refer to this as the weakly-directional optimization. It is important to be able to reject electron backgrounds for the weakly-directional scenario because the analysis threshold for a directional search is expected to be in this energy range. At even lower energies better electron rejection improves the non-directional reach of the experiment (since a directional detector can also operate as a non-directional detector).

\begin{figure}[tbp]
\centering 
\includegraphics[width=.8\textwidth]{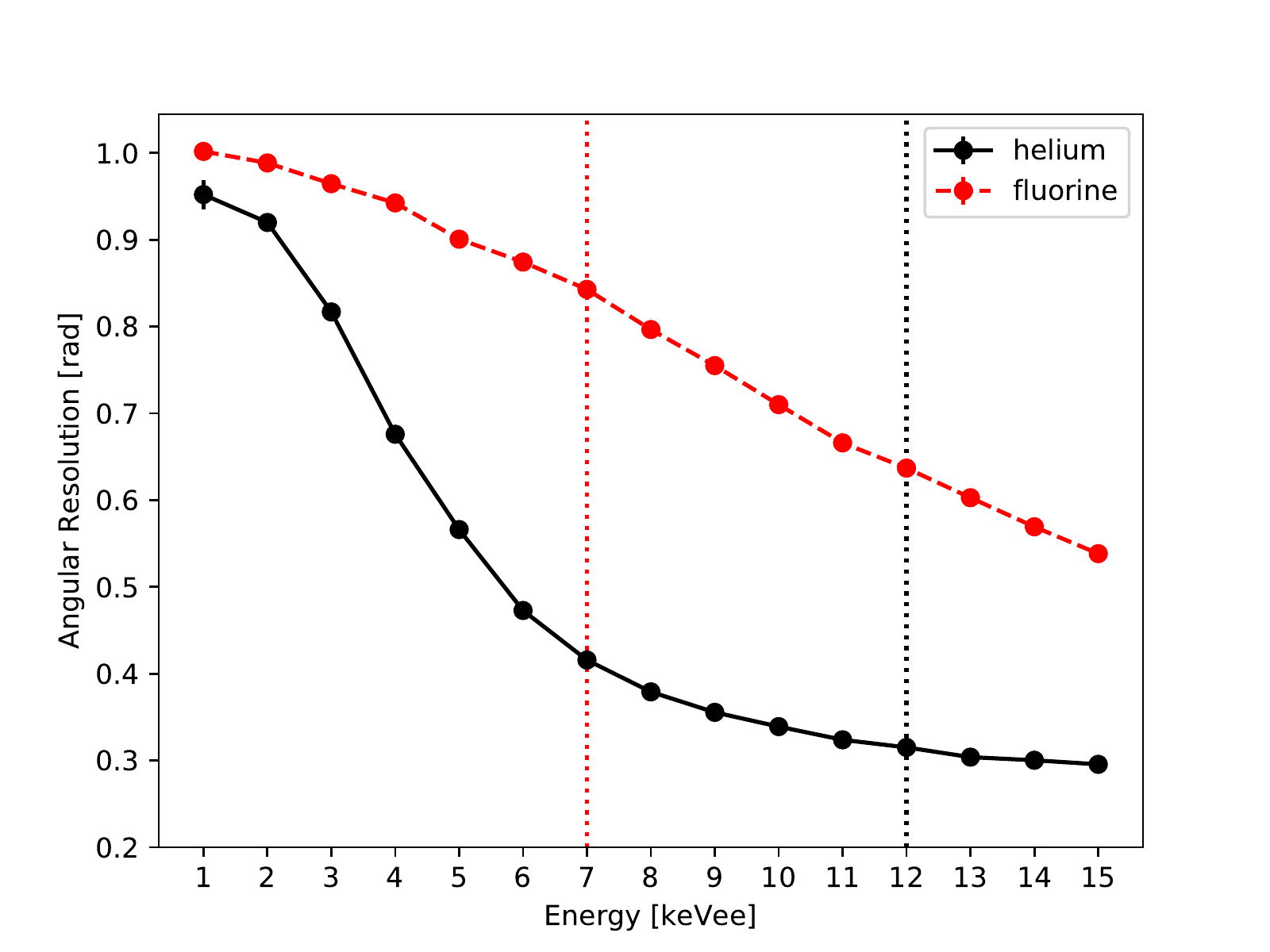}
\hfill

\caption{\label{fig:ang_res} Angular resolution of nuclear recoils versus energy. The vertical line at \SI{7}{\kevee} corresponds to a weakly-directional energy for the fluorine recoils and the vertical line at \SI{12}{\kevee} corresponds to a directional energy for helium recoils. Simulation of recoils assumes a $80 \% \textrm{ He} + 10 \% \textrm{ CF}_4 + 10 \% \textrm{ CHF}_3$ gas mixture at total pressure of $60$ Torr,  a temperature of \SI{25}{\degree C}  and \SI{25}{cm} of drift in a $40.6$ V/cm drift field}
\end{figure}

To determine suitable energies that correspond to directional and weakly-directional scenarios in our gas mixture, we perform the following analysis on our simulated helium and fluorine recoils:
\begin{enumerate}
\item For each recoil use singular value decomposition (SVD) to determine the principal axis.
\item Determine $\Delta \theta$, the angle between the true initial recoil direction and the principal axis.
\item The angular resolution for each energy bin is calculated as the average $\Delta \theta$ over all recoils of a particular species within the bin.
\end{enumerate}
Note that the angle is between two axes, as opposed to two signed vectors. With this convention, the angular resolution is exactly 1 radian in the limit of no directional sensitivity. This is the same procedure as in the \Cygnus feasibility study~\cite{cygnus} and is motivated in ref.~\cite{sven_review}. The angular resolution is plotted over several energy bins for both recoil species in Figure~\ref{fig:ang_res}. We now use Figure~\ref{fig:ang_res} to select suitable cases for the directional and weakly-directional optimization scenarios. For the weakly-directional optimization scenario, we choose to use fluorine recoils in the \SI{7}{\kevee} energy bin as it has an angular resolution close to 1 radian and energy that is large enough to avoid detection threshold limitations. For the directional scenario, we choose helium recoils in the \SI{12}{\kevee} energy bin as their angular resolution is within the plateau seen in Figure~\ref{fig:ang_res} and the energy is low enough for electron rejection to be challenging. A recent review article~\cite{sven_review} identified an angular resolution of 30 degrees (0.52 radians) as a goal for future directional detectors. Our weakly-directional and directional cases are well below and above the 0.52 radian goal, respectively.  An illustration of a recoil track for each of these scenarios is presented in Figure~\ref{fig:tracks}.

\subsection{Length Along Principal Axis (LAPA)}
\label{LAPA}
The first observable that we investigate is the traditional one, used to discriminate electrons in \Cygnus~\cite{cygnus}. It is the projected length of the recoil track along its principal axis, and it is calculated as follows:
\begin{enumerate}
	\item Determine the principal axis of the charge distribution; we use SVD for this step.
	\item Project all of the points in the charge distribution onto the principal axis.
	\item The \texttt{Length Along Principal Axis} (LAPA) is the difference between the maximum and minimum projection values. 
\end{enumerate}
Since this is the traditional observable that we wish to compare against, we do not alter this definition. 
We note that since LAPA is length (for fixed energy) and dE/dX is the average energy divided by the track length, the two observables are equivalent in terms of their ability to distinguish recoils. 

\subsection{Standard Deviation of Charge Distribution (SDCD)}
\label{SDCD}
A limitation of the LAPA definition is its reliance on using SVD to find a principal axis. As the diffusion increases, the principal axis is washed out, which diminishes LAPA's reliability. For this reason, an obvious candidate that is likely to improve performance in the weakly-directional region is the standard deviation of the charge distribution (SDCD)
\begin{equation}
SDCD = \sqrt{ \frac{ \sum_{i=1}^{N} (\mathbf{r_i} - \mathbf{\bar{r}} )^2}{N}}.
\end{equation}
In the above equation, $\mathbf{r_i} $ is the position vector for each (binned) charge, $\mathbf{\bar{r}} $ is the average over $\mathbf{r_i} $, and $N$ is the total number of charges. Since SDCD has no free parameters in its definition, there is no need to optimize it.

\subsection{Clustering Threshold (ClustThres)}
\label{ClustThres}
Another property that has proven to be useful for distinguishing electron recoils from nuclear recoils is the clustering within their charge distributions. Compared with nuclei, electrons follow a more tortuous path through the gas. Since their mass is equal to the electrons that they ionize, large fractions of their energy may be lost in a single interaction. Furthermore, electron-nuclear interactions may occur, which abruptly change the direction of the electron ~\cite{raddect}. This sporadic behavior leaves a charge distribution that is naturally grouped into multiple clusters, unlike nuclear recoils that typically look like a single cluster. We quantify this with the Clustering Threshold (ClustThres). This clustering algorithm utilizes DBScan~\cite{DBScan}, which clusters spatial points using two parameters, min\_samples (the minimum number of neighboring points to be considered a core point) and $\varepsilon$ (the radius of the neighborhood surrounding each point). ClustThres is defined as follows:
\begin{enumerate}
	\item Assign each electron a random position drawn uniformly from within the \SI{100}{\micro m} $\times$ \SI{100}{\micro m} $\times$ \SI{100}{\micro m} readout pixel where it was detected in simulation.
	\item Cluster the charge distribution using the \texttt{DBScan}~\cite{DBScan} with $\varepsilon=  \frac{\textrm{pixel width}} {10}$ and min\_samples = $n$ (where $n$ is a free parameter). 
	\item If the fraction of charge in the largest cluster is less then $thres$ (another free parameter), increment $\varepsilon$ by $d = \frac{\textrm{pixel width}} {10} $ then return to step 2. 
	\item The final value of $\varepsilon$ is defined as ClustThres.
\end{enumerate}

\begin{figure}
    \begin{subfigure}{.49\textwidth}
        \centering
        \SI{12}{\kevee} helium recoils
        \includegraphics[width=\linewidth]{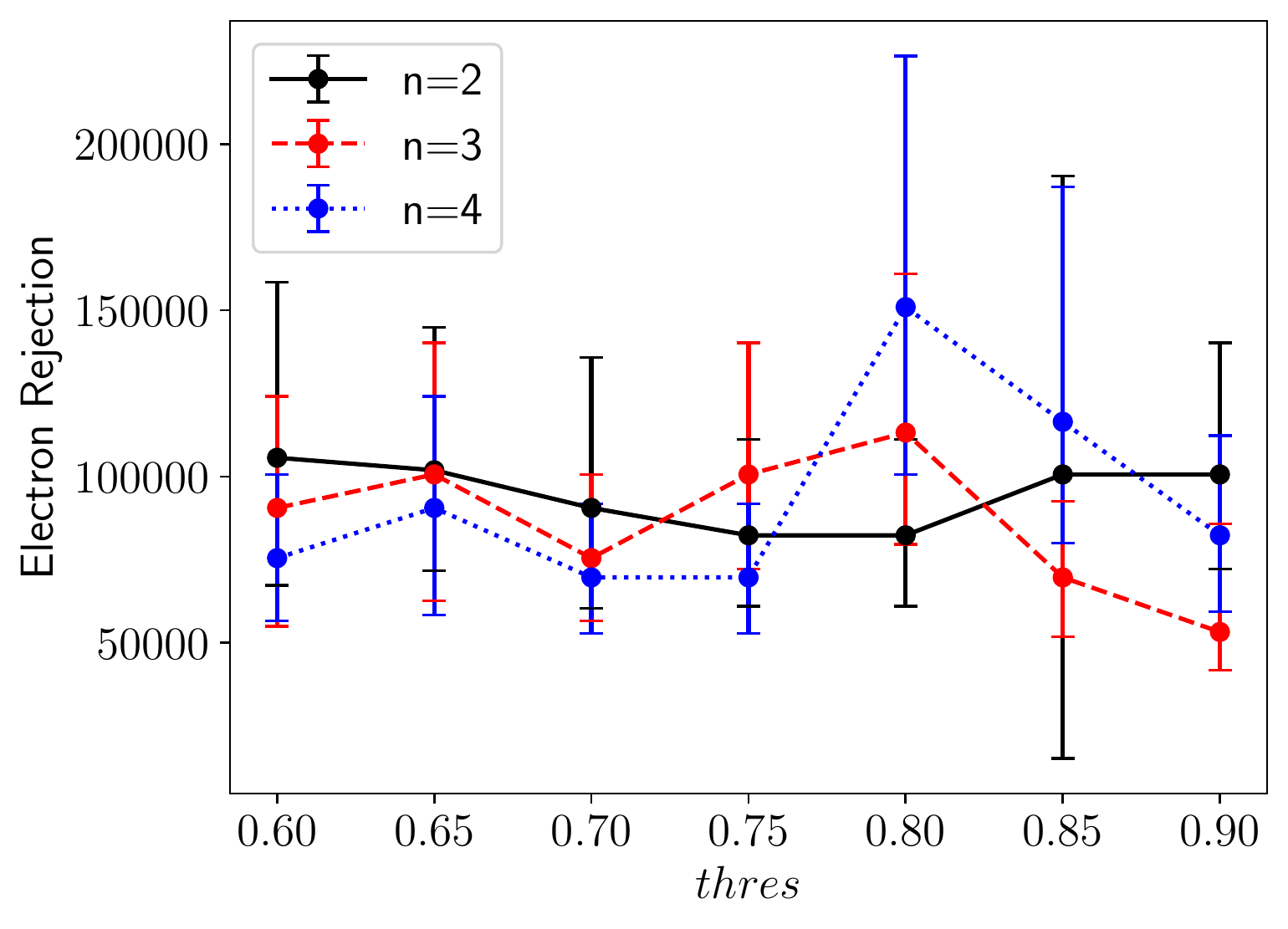}
    \end{subfigure}\hfill
    \begin{subfigure}{.49\textwidth}
        \centering
        \SI{7}{\kevee} fluorine recoils
        \includegraphics[width=\linewidth]{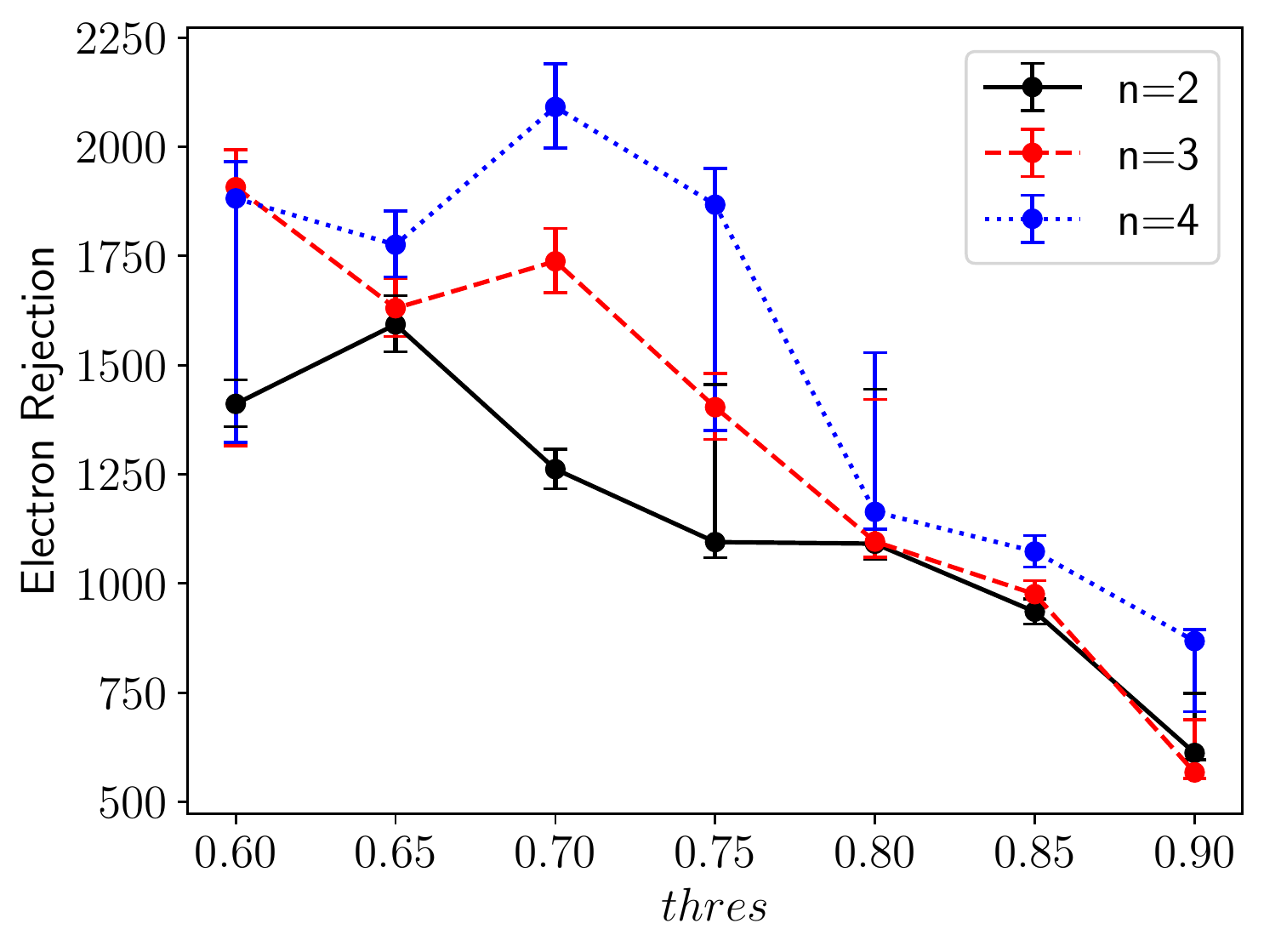}
    \end{subfigure}
    \caption{\label{fig:opt_ClustThres} Electron rejection versus the $thres$ parameter in Section~\ref{ClustThres}, for several integer values of the $n$ parameter. The nuclear recoil selection efficiency is fixed at $\epsilon = 0.5$ for all cases.  Left: helium recoils in the \SI{12}{\kevee} energy bin (the directional case). Right: fluorine recoils in the \SI{7}{\kevee} energy bin (the weakly-directional case). The electron rejection is optimized at $n=4$ and $thres=0.70$ for the weakly-directional case or  $n=4$ and $thres=0.80$ for the directional case. }
\end{figure}

In the above definition, both $n$ and $thres$ are free parameters; in order to specify them, we demand that they optimize electron rejection. We conduct separate optimizations for the directional and weakly-directional cases discussed at the beginning of Section~\ref{sec:define}. To optimize the parameters, we fix the nuclear recoil selection efficiency at $\epsilon = 0.5$ and assess electron rejection over a range of the parameters' values. In Figure~\ref{fig:opt_ClustThres} we see that for the weakly-directional case the electron rejection is optimized when $n=4$ and $thres=0.70$, whereas for the directional case the optimal values are $n=4$ and $thres=0.80$. We note that the maxima for the electron rejection in Figure~\ref{fig:opt_ClustThres} have relatively low statistical significance. Hence, we do not expect the choice of $n$ and $thres$ in ClustThres to have a significant effect on electron rejection. 

\subsection{Number of Clusters (NumClust)}
\label{NumClust}
\begin{figure}
    \begin{subfigure}{.49\textwidth}
        \centering
        \SI{12}{\kevee} helium recoils
        \includegraphics[width=\linewidth]{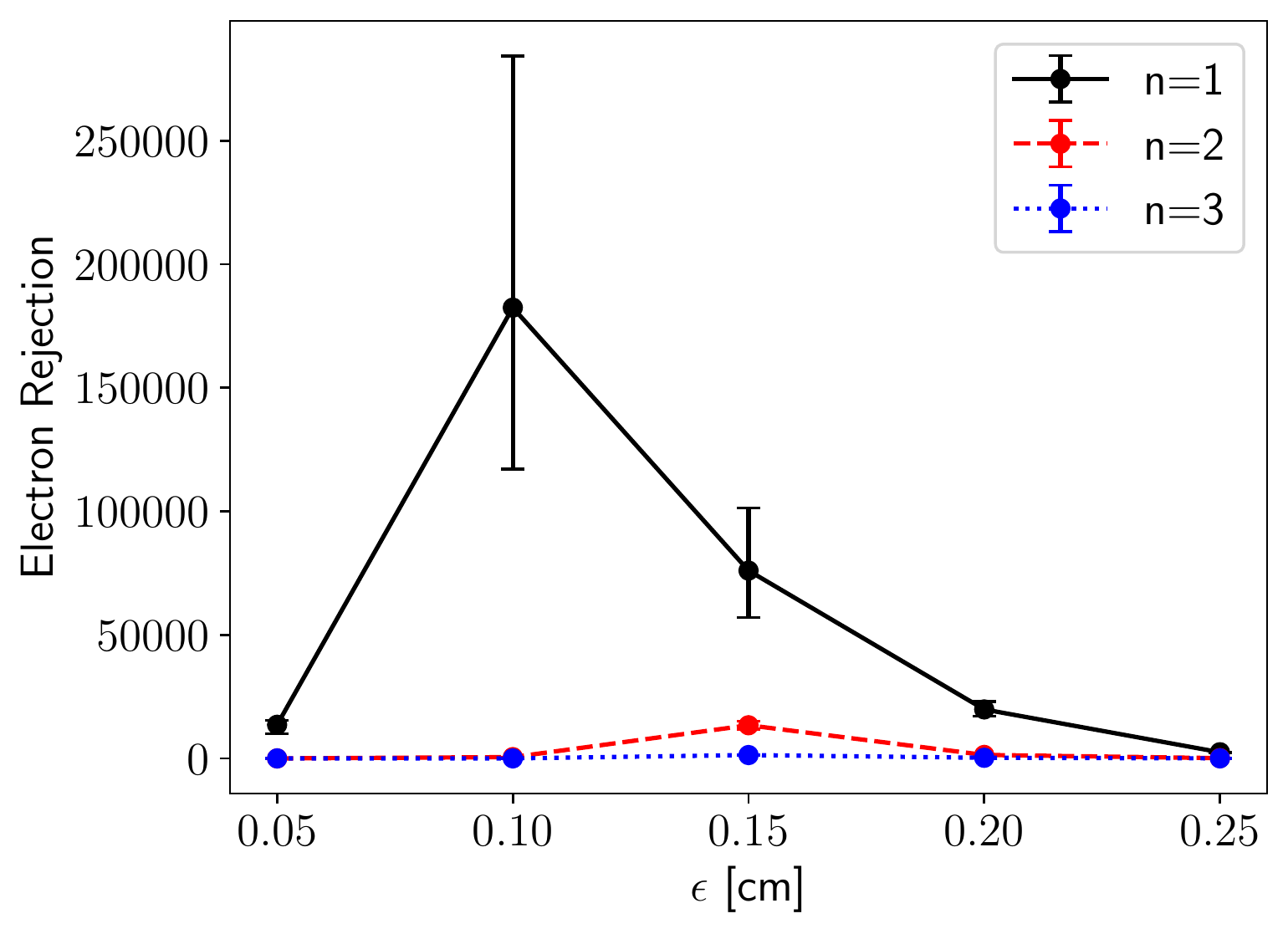}
    \end{subfigure}\hfill
    \begin{subfigure}{.49\textwidth}
        \centering
        \SI{7}{\kevee} fluorine recoils
        \includegraphics[width=\linewidth]{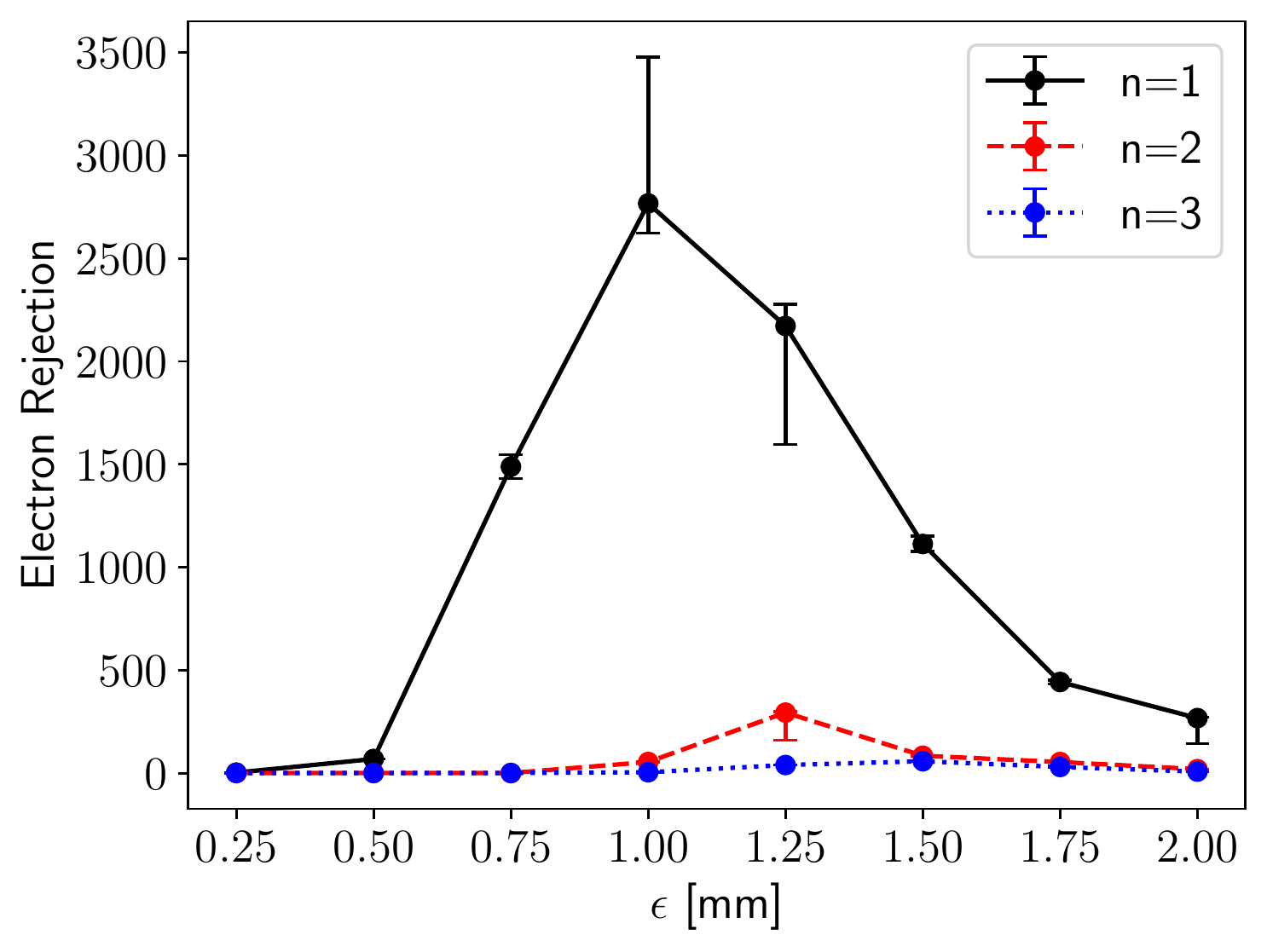}
    \end{subfigure}
    \caption{\label{fig:opt_NumClust} Electron rejection versus the $\epsilon$ parameter in Section~\ref{NumClust}, for several integer values of the $n$ parameter. The nuclear recoil selection efficiency is fixed at $\epsilon = 0.5$ for all cases. Left: helium recoils in the \SI{12}{\kevee} energy bin (the directional case). Right: fluorine recoils in the \SI{7}{\kevee} energy bin (the weakly-directional case). For both directional and weakly-directional cases, the optimal parameter values are $n=1$ and $\varepsilon =\SI{0.1}{cm}$. }
\end{figure}

Another way to quantify the behavior described in Section~\ref{ClustThres} is to count the number of clusters. We again use the \texttt{DBScan}~\cite{DBScan} algorithm, except this time, our free parameters are $\varepsilon$ and $n$. Once again, we specify both parameters by demanding that they optimize electron rejection while fixing the nuclear recoils selection efficiency at $\epsilon = 0.5$, for the directional and weakly-directional cases separately. In Figure~\ref{fig:opt_NumClust}, we see that for both directional and weakly-directional cases, the optimal parameter values are $n=1$ and $\varepsilon =  \SI{0.1}{cm} $. Although we arrived at the same optimization for both cases, Figure~\ref{fig:opt_NumClust} demonstrates that it is important to optimize this observable.

\subsection{Maximum Density (MaxDen)}
\label{MaxDen}

\begin{figure}
    \begin{subfigure}{.49\textwidth}
        \centering
        \SI{12}{\kevee} helium recoils
        \includegraphics[width=\linewidth]{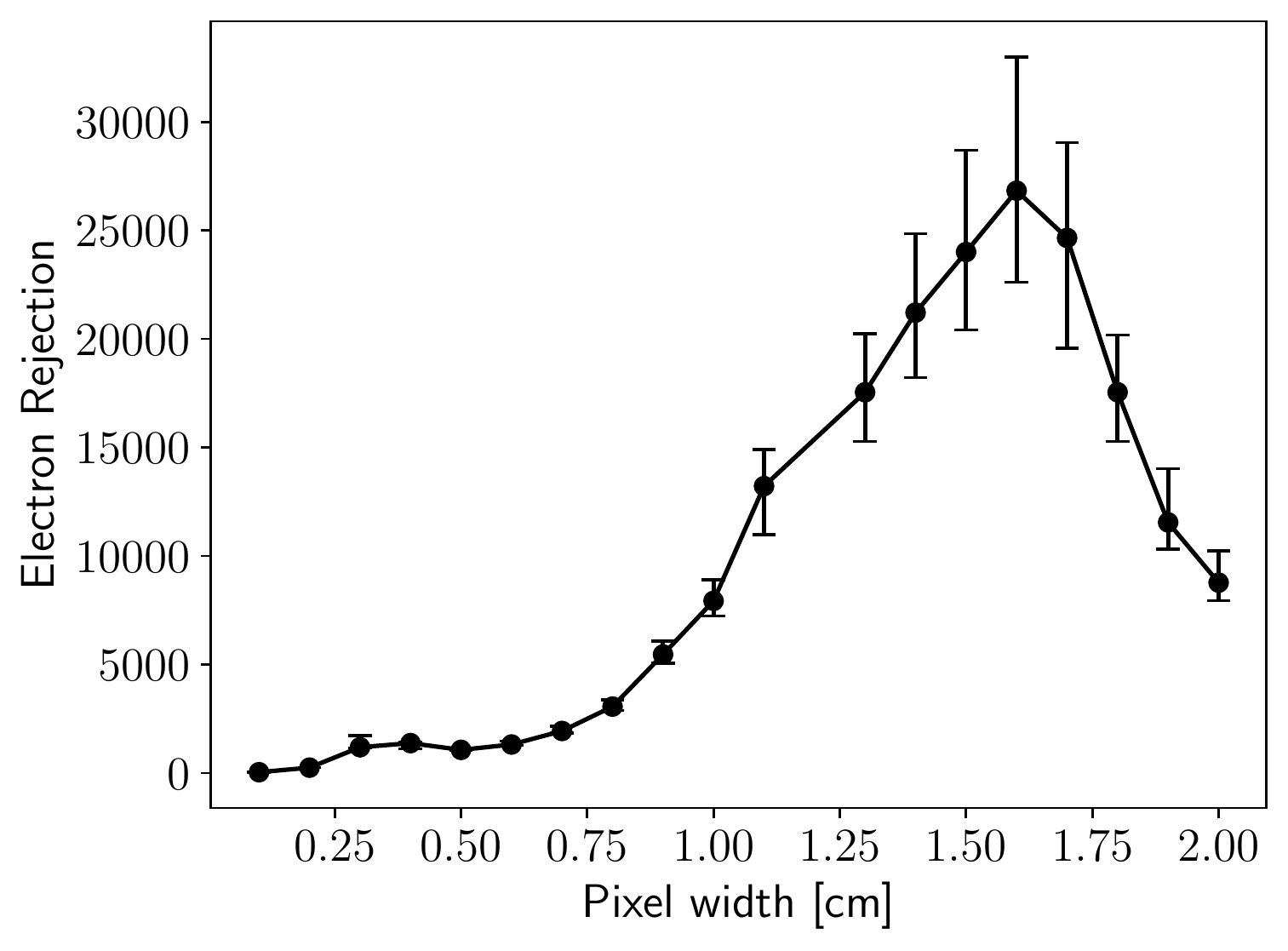}
    \end{subfigure}\hfill
    \begin{subfigure}{.49\textwidth}
        \centering
        \SI{7}{\kevee} fluorine recoils
        \includegraphics[width=\linewidth]{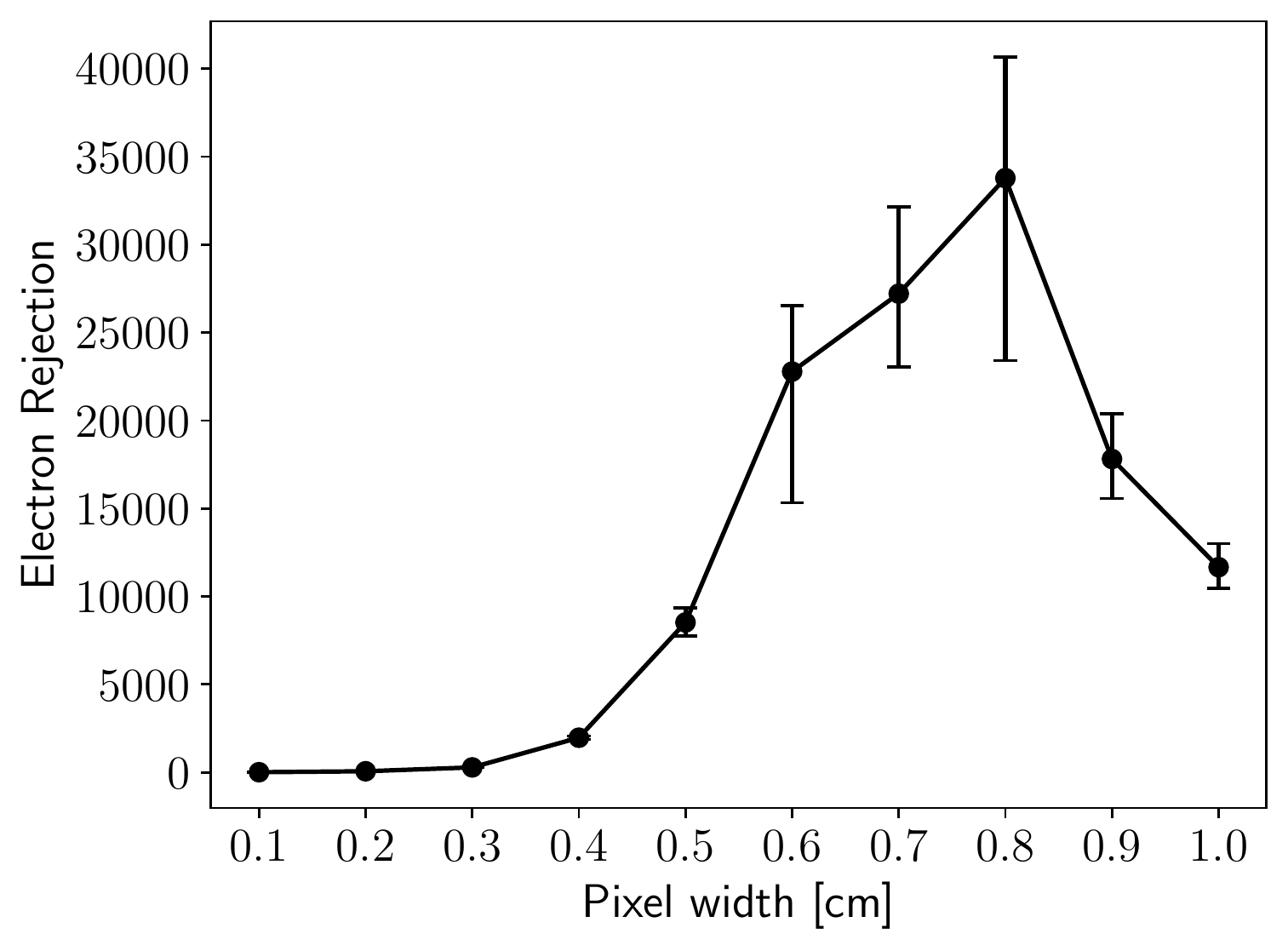}
    \end{subfigure}
    \caption{\label{fig:opt_MaxDen} Electron rejection versus the pixel width parameter in Section~\ref{MaxDen}. The nuclear recoil selection efficiency is fixed at $\epsilon = 0.5$ for all cases.   Left: helium recoils in the \SI{12}{\kevee} energy bin (the directional case). Right: fluorine recoils in the \SI{7}{\kevee} energy bin (the weakly-directional case). The optimal pixel width for the weakly-directional scenario is  \SI{0.8}{cm} and for the directional scenario it is  \SI{1.6}{cm}.}
\end{figure}

Electrons lose their energy at a slower rate than nuclei, even though an electron can lose a larger fraction of its energy per interaction. This suggests that electron recoils travel a greater distance between interactions resulting in more sparse distributions. A simple measure of this is the maximum density within the charge distribution (MaxDen). We choose to calculate MaxDen as the inverse of the maximum number of electrons within a pixel. The recoil tracks are already binned into $100$ $\si\micro$m $\times$ $100$ $\si\micro$m $\times$ $100$ $\si\micro$m pixels. To optimize MaxDen for directional and weakly-directional recoils, we choose to re-bin the tracks into larger pixels of varying width. With pixel width as the free parameter, we optimize Maximum Density for the directional and weakly-directional case exactly as we have done for the previous two observables. In Figure~\ref{fig:opt_MaxDen} the electron rejection is plotted against pixel width. We see that the optimal bin size for directional recoils is \SI{1.6}{cm} and the optimal bin size for weakly-directional recoils is \SI{0.8}{cm}. For this observable, we see prominent and unique optimizations for the directional case and the weakly-directional case.

\subsection{Charge Uniformity (ChargeUnif)}
\label{ChargeUnif}

While electron recoils tend to have charge distributions that are dense in some areas and sparse in others, nuclear recoils are generally uniform. This motivates another direction-independent observable, the Charge Uniformity (ChargeUnif). This observable is a measure of how uniform a recoil track is, and it is defined as follows:
\begin{enumerate}
	\item For each point within the charge distribution, find the average distance to all other points.
	\item The \texttt{ChargeUnif} is the standard deviation of the values calculated in step 1.
\end{enumerate}
In the definition above, there are no free parameters and so optimization is not required. 

\subsection{Cylindrical Thickness (CylThick)}
\label{CylThick}
LAPA finds a principal axis of a recoil track and then only uses information along that axis. An alternative approach is to use the information on the plane perpendicular to the principal axis. As electrons travel through the gas, they experience far more pronounced straggling than the nuclei.  This means that the principal axis approximates a nuclear recoil's trajectory much more accurately than it does for an electron recoil. For this reason, we calculate the Cylindrical Thickness (CylThick) as
\begin{enumerate}
	\item Find the principal axis of the charge distribution.
	\item For each charge, calculate the squared distance to the principal axis.
	\item CylThick is the sum of all squared distances.
\end{enumerate}
CylThick is interpreted as measure of how much a recoil track deviates from the trajectory approximated by the principal axis. This observable is fully defined and does not need to be optimized.

\section{Results}
\label{sec:results}
Now that we have defined our observables, we may assess their electron rejection using the methodology discussed in Section~\ref{assessER}. In Section~\ref{Non-directional} we assess the electron rejection for the weakly-directional optimization scenario, analyzing F and $e^-$ recoils in the \SI{7}{\kevee} energy bin. In Section~\ref{directional} we assess electron rejection for the directional optimization scenario, analyzing He and $e^-$ recoils in the \SI{12}{\kevee} energy bin. Although we have only investigated two optimizations, we show the robustness of our observables and their optimizations by investigating electron rejection over a range of energies in Section~\ref{other}.

\subsection{Weakly-Directional Recoils}
\label{Non-directional}

\begin{figure}[tbp]
\centering 
\includegraphics[width=.8\textwidth]{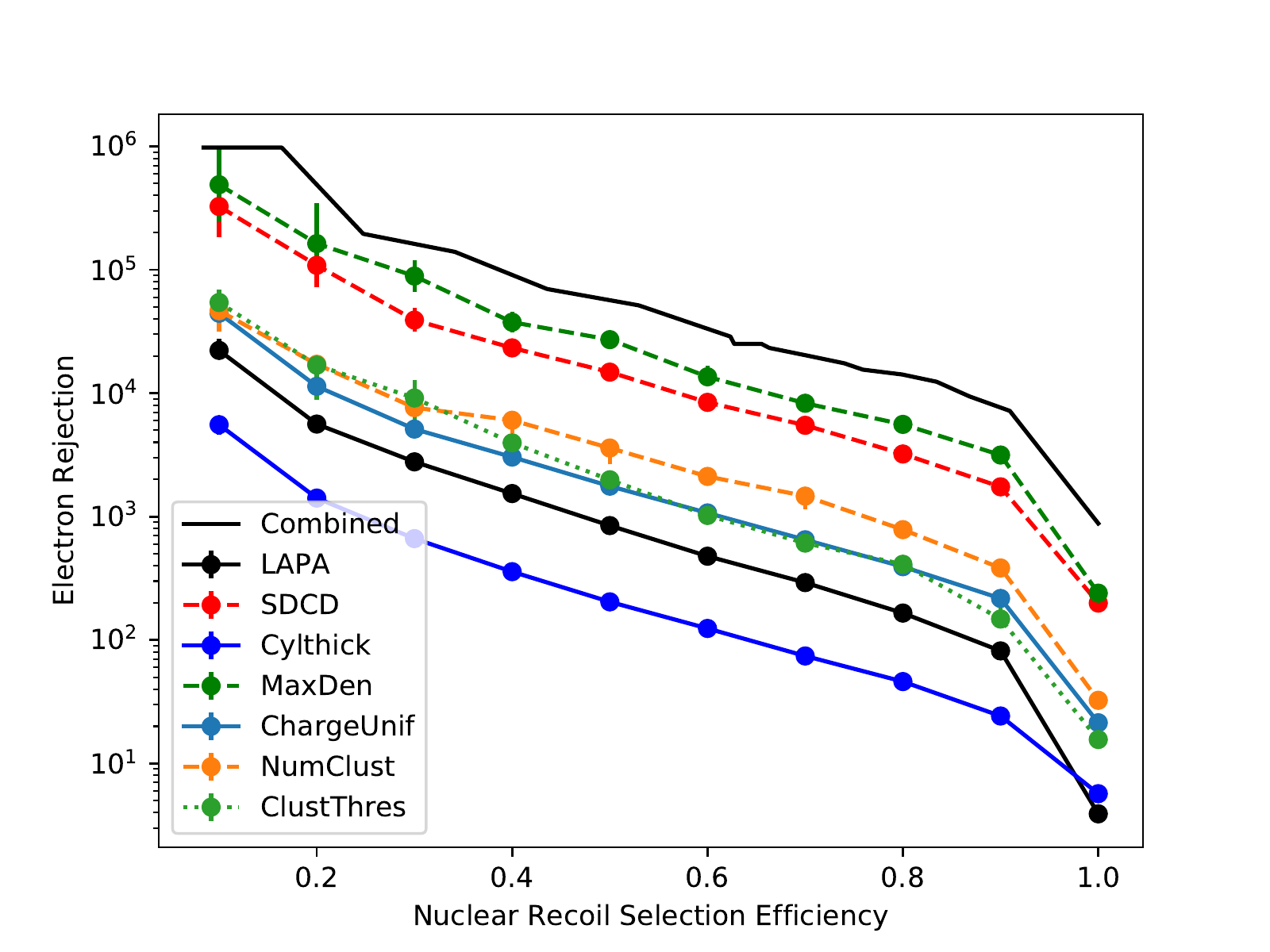}
\hfill
\caption{\label{fig:result_nondir} Electron Rejection versus Nuclear Recoil Selection Efficiency for all observables. For NumClust, ClustThres, and MaxDen we use the definitions that we optimized for the weakly-directional case in Section~\ref{sec:define}. These results are for the fluorine and $e^-$ recoil simulations in the \SI{7}{\kevee} energy bin.}
\end{figure}

To assess electron rejection for the \SI{7}{\kevee} fluorine and $e^-$ recoils, we calculate the electron rejection factor $(R)$ corresponding to a specified nuclear recoil efficiency $(\epsilon)$, as described in Section~\ref{assessER}. In Figure~\ref{fig:result_nondir}, we plot $R$ versus $\epsilon$ for all observables calculated on the weakly-directional simulation set containing fluorine and electron recoils in the \SI{7}{\kevee} energy bin. We note that all observables except for CylThick outperform LAPA. MaxDen and SDCD achieve the highest electron rejection for the weakly-directional case.

\begin{figure}[tbp]
\centering 
\includegraphics[width=.49\textwidth]{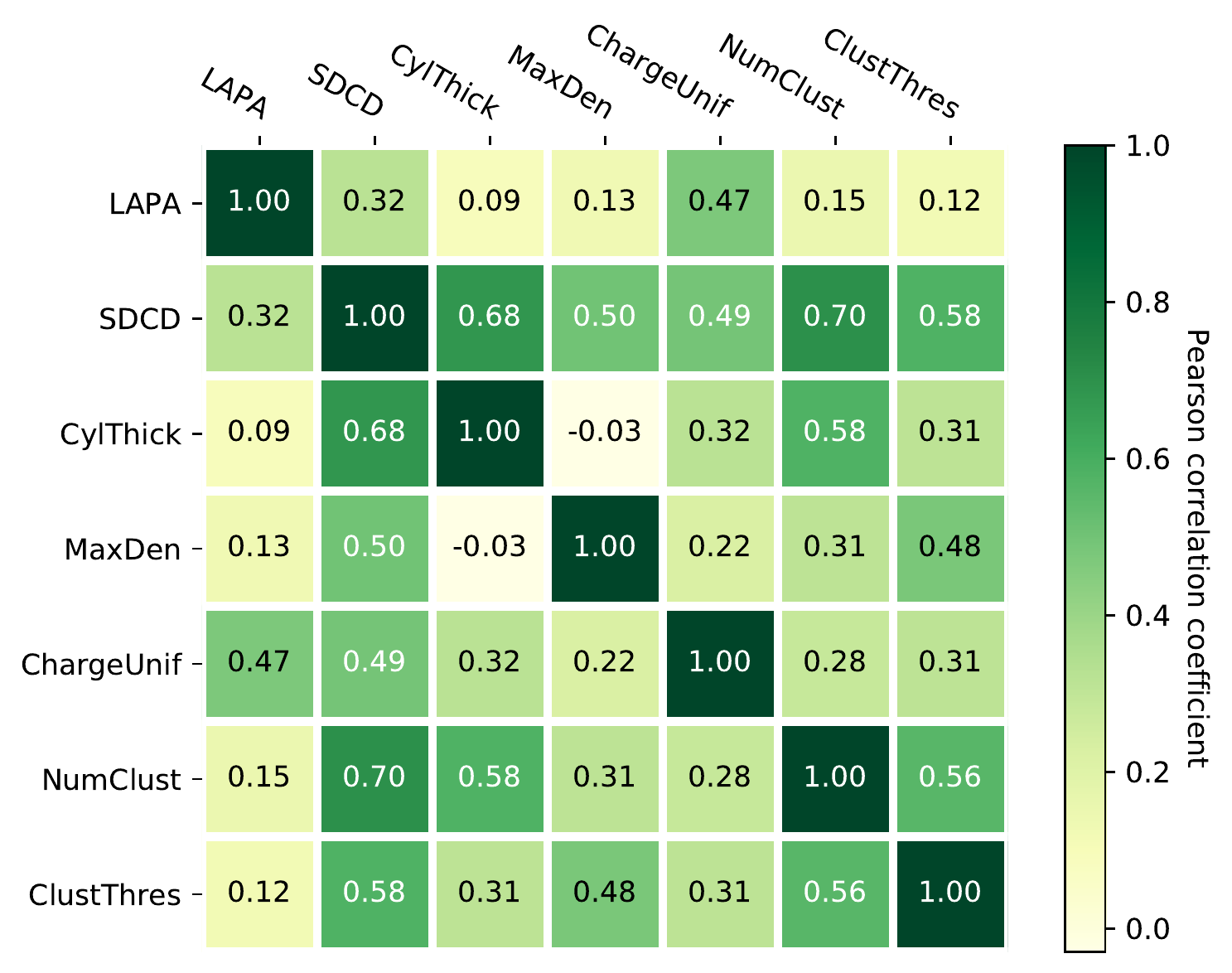}
\hfill
\includegraphics[width=.49\textwidth]{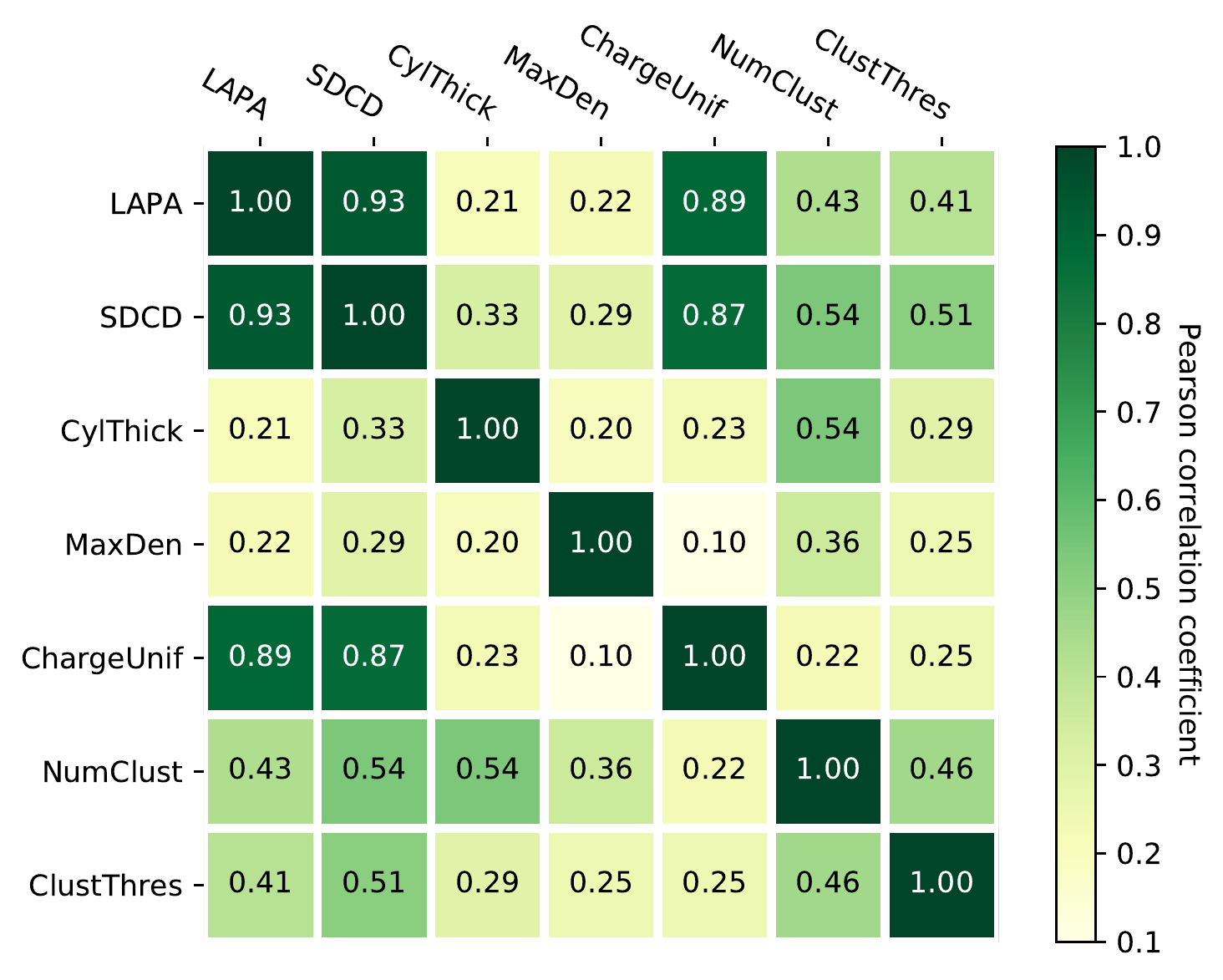}
\caption{\label{fig:corr_7} The Pearson product-moment correlation coefficients between all observables. The left figure shows the correlations for the fluorine recoils in the \SI{7}{\kevee} energy bin and the right figure shows the correlations for the $e^-$ recoils in the \SI{7}{\kevee} energy bin.}
\end{figure}

To further improve electron rejection, it would be appealing to use several of these observables simultaneously. To see how effectively these observables can be combined, Figure~\ref{fig:corr_7} displays the Pearson correlation coefficients of the observables. For the electron recoils, we see strong correlations between LAPA, SDCD, and ChargeUnif. Of these three observables, it is clear that for the weakly-directional case, SDCD is the most valuable. The remaining observables are weakly correlated with one another, suggesting that the electron rejection can be improved significantly with a combined observable.

\begin{figure}
     \centering
     \begin{subfigure}[b]{0.49\textwidth}
         \centering
         \includegraphics[width=\textwidth]{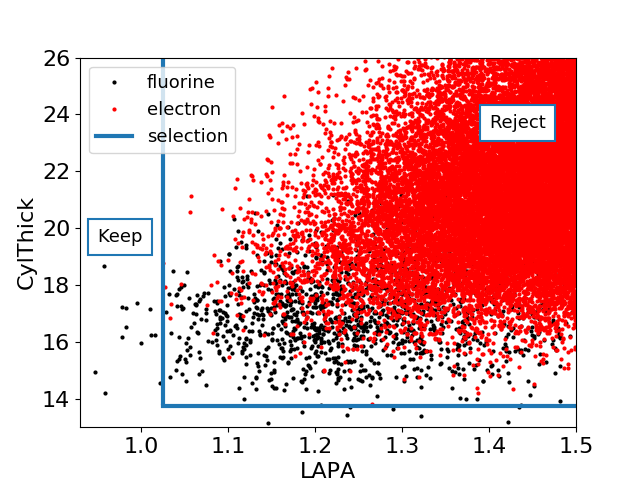}
         \caption{Pre-selection criteria: Rejected recoils are passed on to Method 1 or 2.}
         \label{a}
     \end{subfigure}
     \centering
     \hfill
     \begin{subfigure}[b]{0.49\textwidth}
         \includegraphics[width=\textwidth]{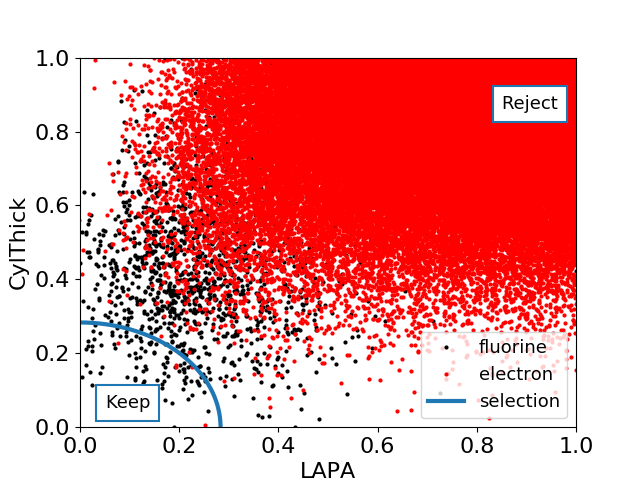}
         \caption{Method 1: Spheres are used to separate the recoil species.}
         \label{b}
     \end{subfigure}\\
     \hfill
     \begin{subfigure}[b]{0.49\textwidth}
         \includegraphics[width=\textwidth]{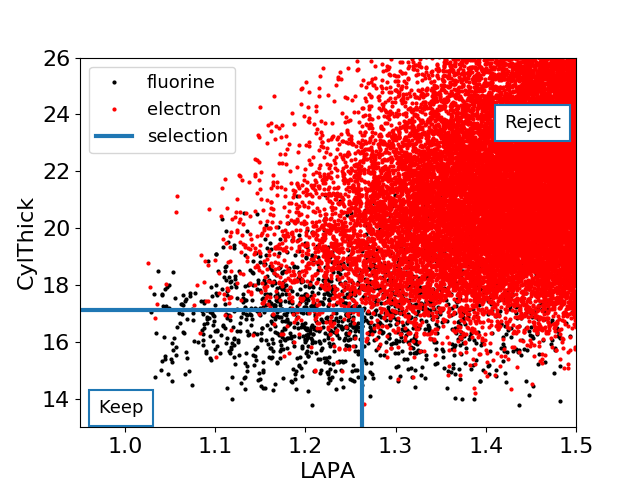}
         \caption{Method 2: Rectangles are used to separate the recoil species.}
         \label{c}
     \end{subfigure}
     \hfill
     \begin{subfigure}[b]{0.49\textwidth}
         \includegraphics[width=\textwidth]{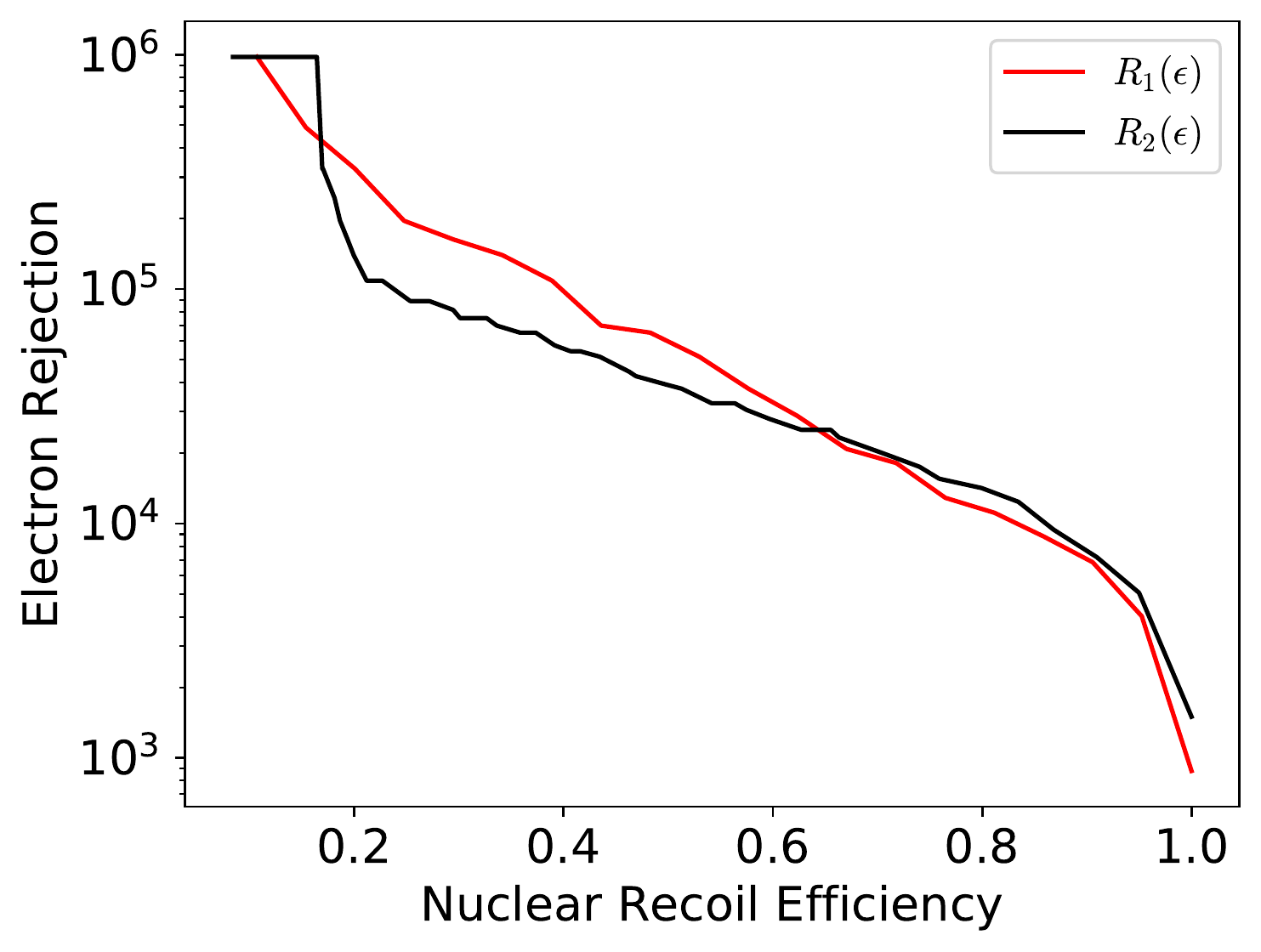}
         \caption{Electron Rejection versus Nuclear Recoil Efficiency plots obtained from Methods 1 and 2.}
         \label{d}
     \end{subfigure}
        \caption{ An illustration of how the combined observable is computed. Subfigures~\ref{a}--\ref{c} are projections that only show two of the seven observables used in the calculation of Subfigure~\ref{d}. Subfigure~\ref{a} displays an example of our pre-selection criteria, the recoils rejected by this step are passed on to either separation Method 1 or 2, shown in Subfigures~\ref{b} and~\ref{c}, respectively. In Method 1, the spherical selection criteria is used to separate the fluorine recoils from the electron recoils. In Method 2, the rectangular selection criteria is used to separate the two recoils. Subfigure~\ref{d} displays the electron rejection factor obtained by the pre-selection criteria followed by Method 1 or Method 2. The combined observable is defined as $\textrm{max}(R_1(\epsilon),R_2(\epsilon))$.}
        \label{fig:three graphs}
\end{figure}

Here, we adopt a simple approach for combining observables, and we note that it can be improved in a dedicated study. All observables are defined to have lower values for nuclear recoils than for electron recoils. As a result, in observable space, nuclear recoils will typically be closer to the origin, as shown for the CylThick and LAPA observables in Figure~\ref{a}. With recoils plotted in observable space, it is straightforward to impose geometric selection criteria to separate the electron recoils from the nuclear recoils. Our first step is to impose pre-selection criteria, designed to keep as many nuclear recoils as possible before we attempt to separate the two recoil species. Along each axis in observable space, we find the minimum value for the electron recoil distribution and keep all recoils below this value. By imposing this pre-selection criteria, we first keep nuclear recoils that can be clearly identified with a single observable (even though they may not be clearly identified by other observables). This step allows the recoil separation process to begin at a higher nuclear recoil efficiency, improving electron rejection at low nuclear recoil efficiency. Figure~\ref{a} shows an illustration of this pre-selection step where only the CylThick and LAPA axis have been plotted. The recoils that are rejected by the pre-selection criteria are passed along to two alternative methods that separate the recoil species. The pre-selection efficiency is taken into account in the final evaluation of the nuclear recoil efficiency.

In Method 1, spheres are used to separate the two recoil species in observable space. Since the scales of our observables vary significantly, we shift each axis so that the nuclear recoil distribution starts at 0 and ends at 1. Now we draw origin-centered spheres in observable space; recoils residing within the sphere are kept and recoils outside are rejected, as shown in Figure~\ref{b}. As the radius of the sphere is incremented, more nuclear recoils are kept and the nuclear recoil selection efficiency ($\epsilon$) increases. For each value of $\epsilon$, the corresponding $R$ is calculated until the sphere is large enough to contain all nuclear recoils ($\epsilon = 1$). This approach gives an electron rejection versus nuclear recoil selection efficiency plot denoted $R_1(\epsilon)$ and shown in Figure~\ref{d}. Method 2 is similar to Method 1, except we use rectangles to separate the recoil species, see Figure~\ref{c}. This time, we do not shift or scale any of the axes. As in Method 1, the rectangle must be incremented until all nuclear recoils are contained, to obtain an electron rejection versus nuclear recoil selection efficiency plot (denoted $R_2(\epsilon)$ in Figure 9~\ref{d}). We increment all sides of the rectangle simultaneously, in a manner that is consistent with the scale of the data along its axis. Hence, the increments are taken as steps through the percentile values of the nuclear recoil distribution along each axis (such that the 100th step will contain $100\%$ of nuclear recoils, corresponding to $\epsilon =1.0$). These two methods yield two plots of electron rejection versus nuclear recoil efficiency, they are plotted alongside each other in Figure 9d. We define the combined observable as $\textrm{max}(R_1(\epsilon),R_2(\epsilon))$.

We emphasize that this simple method is only a lower limit of what can be achieved with a more sophisticated method of combining our observables. The combined observable for the weakly-directional case is plotted as the solid black line in Figure~\ref{fig:result_nondir}. Comparing the combined observable to LAPA, we see an improvement of roughly two orders of magnitude.

\subsection{Directional Recoils}
\label{directional} 

\begin{figure}[tbp]
\centering 
\includegraphics[width=.8\textwidth]{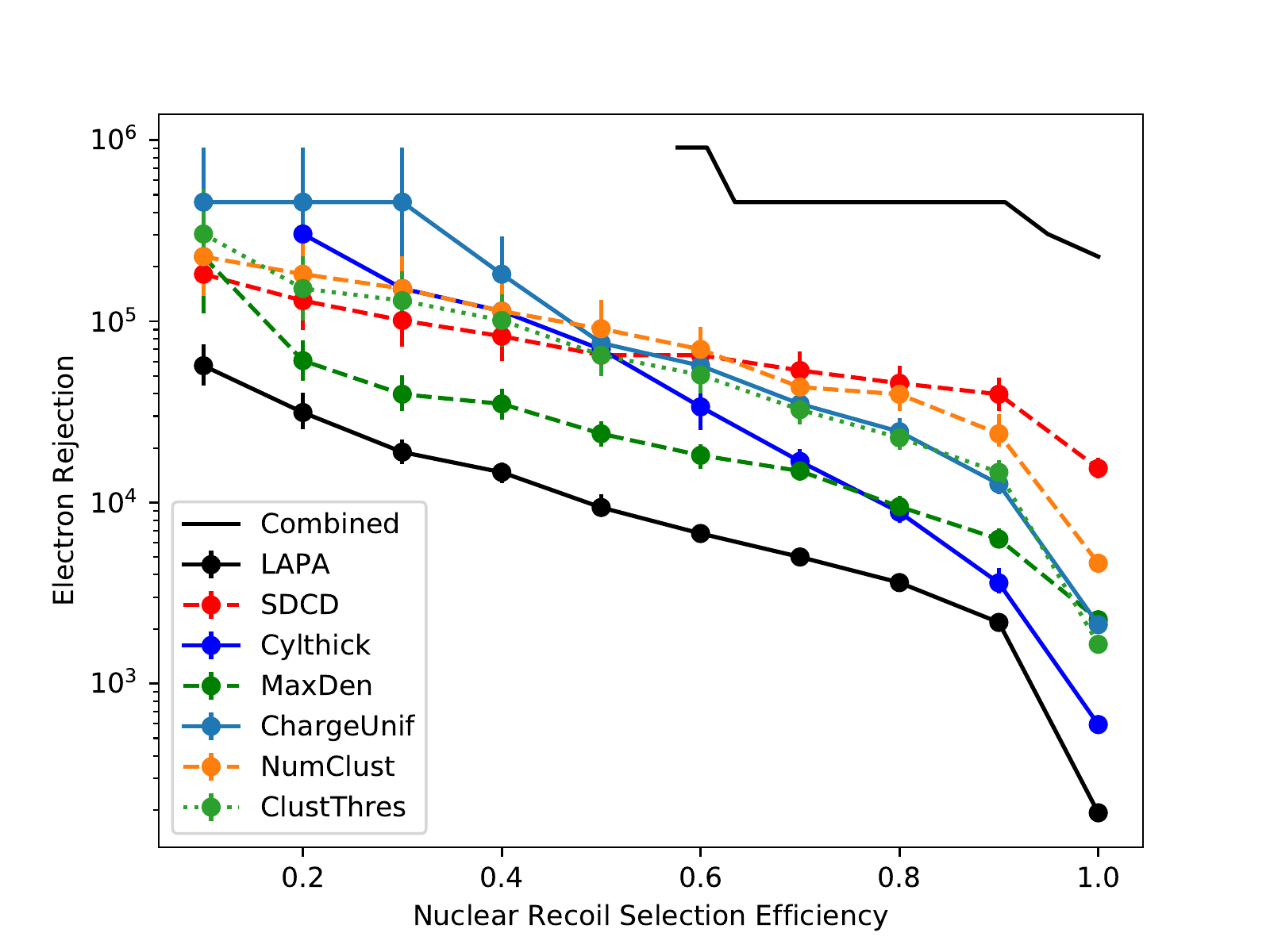}
\hfill
\caption{\label{fig:result_dir} Electron Rejection versus nuclear recoil selection efficiency for all observables. For NumClust, ClustThres, and MaxDen we use the definitions that we optimized for the directional case in Section~\ref{sec:define}. These results are for the helium and $e^-$ recoils in the \SI{12}{\kevee} energy bin. For recoil selection efficiency values where no electron rejection value is shown, it means that the corresponding observable has rejected all simulated electrons.}
\end{figure}

As done in Section~\ref{Non-directional}, we calculate the electron rejection for all observables, this time for the directional scenario with helium and electron recoils in the \SI{12}{\kevee} energy bin. The results are shown in Figure~\ref{fig:result_dir}. If all electrons are rejected for some nuclear recoil selection efficiency, the calculated electron rejection will be infinite, while determining the true electron rejection would require even larger simulation statististics. Our convention is to not show a result in those cases. A missing line in Figure~\ref{fig:result_dir} (i.e. CylThick at 0.1 recoil selection efficiency) thus indicates that the observable has rejected all simulated electrons. For the directional case, all observables outperform LAPA. At low nuclear recoil selection efficiency, CylThick has the best performance rejecting the entire sample of roughly one million electron recoils up until $\epsilon = 0.2$. ChargeUnif also has excellent performance for low nuclear recoil selection efficiency. At higher nuclear recoil selection efficiencies, SDCD and NumClust are the best performers.
 
\begin{figure}[tbp]
\centering 
\includegraphics[width=.49\textwidth]{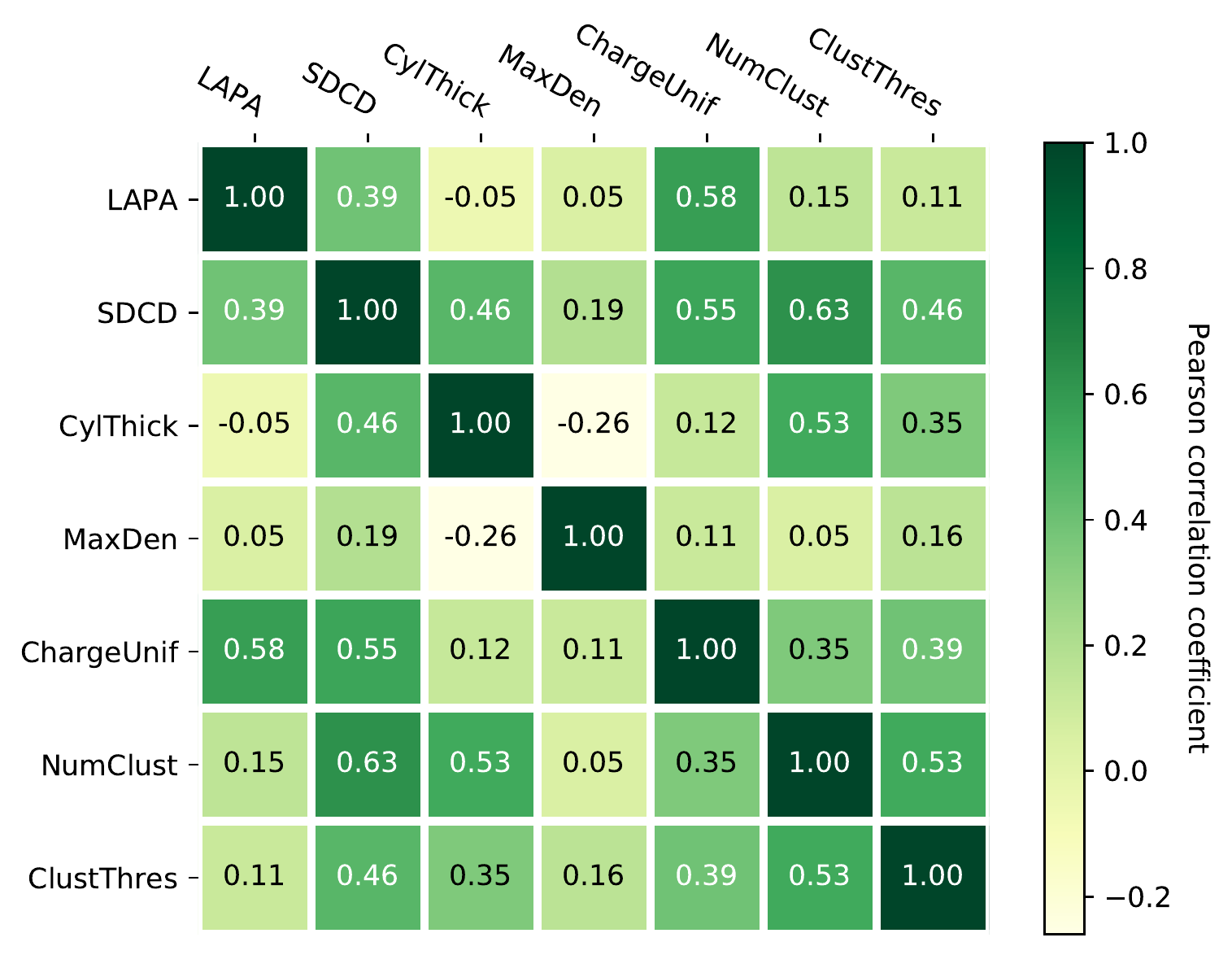}
\hfill
\includegraphics[width=.49\textwidth]{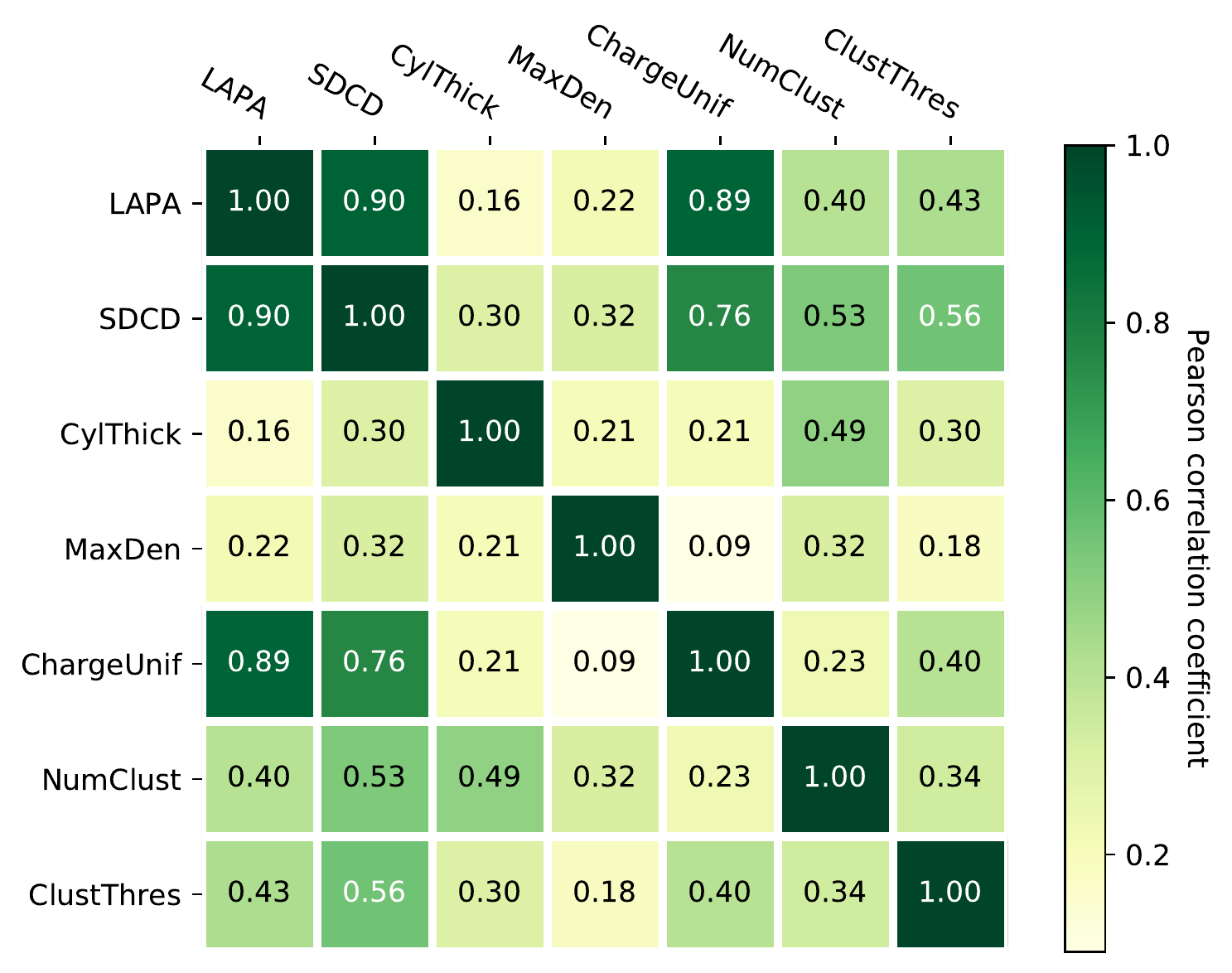}
\caption{\label{fig:corr_12} The Pearson product-moment correlation coefficients between all observables. The left figure shows the correlations for the helium recoils in the \SI{12}{\kevee} energy bin and the right figure shows the correlations for the $e^-$ recoils in the \SI{12}{\kevee} energy bin.}
\end{figure}

The Pearson correlation coefficients of the electron and nuclear recoils in the directional simulation set are shown in Figure~\ref{fig:corr_12}. As concluded in section~\ref{Non-directional}, LAPA, SDCD, and ChargeUnif are strongly correlated while all others are not. Once again, we compute a combined observable for the directional recoils (also plotted  in Figure~\ref{fig:result_dir}). We see a remarkable increase in performance for the directional case when comparing LAPA to the combined observable. The combined observable rejects all (roughly one million) electrons in the \SI{12}{\kevee} energy bin, up to a nuclear recoil selection efficiency of $0.58$. The combined observable improves upon LAPA's electron rejection by two to three orders of magnitude depending on the desired nuclear recoil selection efficiency.

\subsection{Electron Rejection Versus Energy}
\label{other}

To verify that our observables (with both directional and weakly-directional optimizations) are robust, we calculate the electron rejection achieved by each optimized observable over a range of energies (\SI{3}{\kevee} to \SI{15}{\kevee}). To do this, we fix the nuclear recoil selection efficiency at $\epsilon = 0.5$ and calculate the corresponding electron rejection for the electron and nuclear recoil simulations in each energy bin. This is exactly the procedure described in Section 2.2, except for the fact that we fix $\epsilon = 0.5$ in this analysis. Since electron rejection is generally exponential versus energy, we expect to reach an energy bin where all simulated electron recoils are rejected. In this region, we will have reached the limit of our statistics and we can only set a lower limit on $R$. Once all electrons are rejected by an observable, we no longer plot the electron rejection and we terminate the calculation for all of the following energy bins.

\begin{figure}
    \begin{subfigure}{\textwidth}
        \centering
        \includegraphics[width=0.8\linewidth]{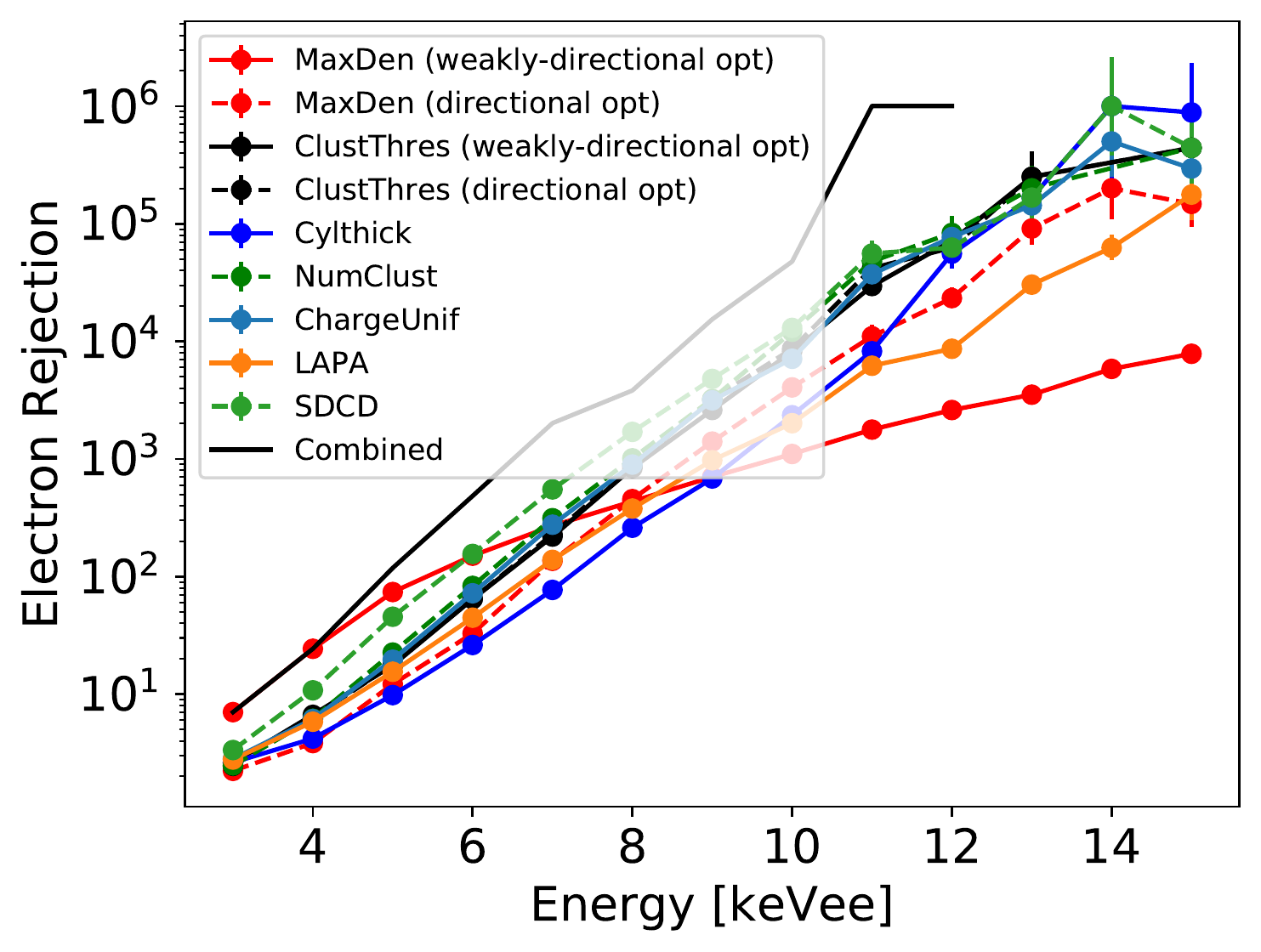}
        \caption{\label{he_rec}helium recoils}
    \end{subfigure}\hfill \\
    \begin{subfigure}{\textwidth}
        \centering
        \includegraphics[width=0.8\linewidth]{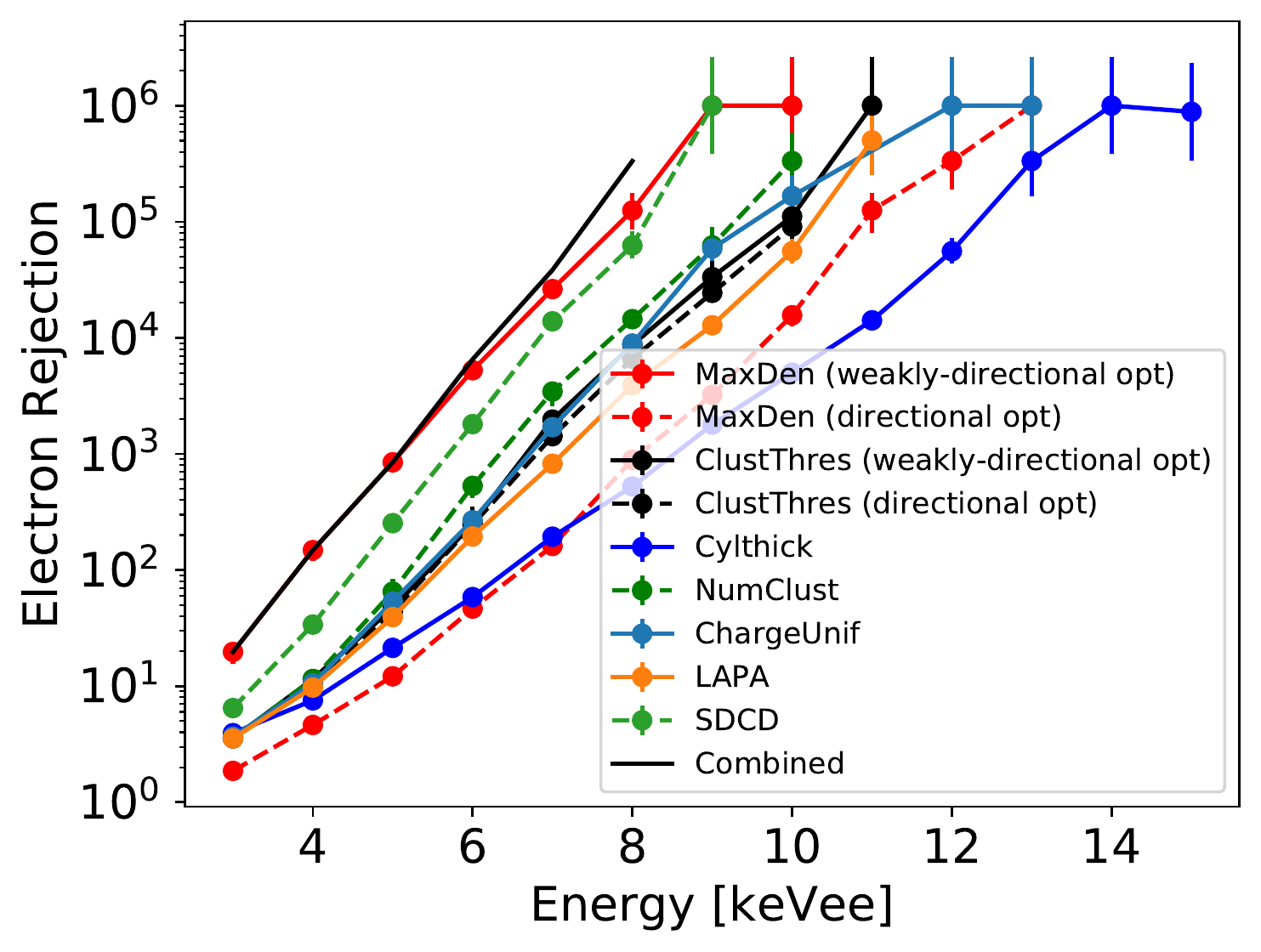}
        \caption{\label{f_rec}fluorine recoils}
    \end{subfigure}
    \caption{\label{fig:all_E} Electron rejection versus electron-equivalent energy for all observables. If an observable is optimized, the optimization case is shown in parenthesis. To calculate the electron rejection, the nuclear recoil selection efficiency was fixed at $\epsilon = 0.5$.   The electron rejection achieved using fluorine recoils is shown on the bottom and the plot on the top is for helium recoils. Once all electrons in the simulation set are rejected by an observable, we no longer plot the electron rejection. The combined observable rejects all (roughly one million) electron recoils starting with the \SI{13}{\kevee} and \SI{9}{\kevee} energy bins for helium and fluorine recoils, respectively.}
\end{figure}

In Figure~\ref{fig:all_E}, we plot electron rejection versus energy for all observables. If an observable has parameters that were optimized, we specify which optimization was used in parenthesis in the legend. One observation is that our observables were able to reject all simulated electrons at a lower energy for fluorine recoils than for helium recoils. This behavior is expected because helium will have a longer, more electron-like recoil track. 

For the fluorine recoils plot in Figure~\ref{f_rec}, the weakly-directional optimization of MaxDen has the best performance, whereas the directional optimization of MaxDen and Cylthick have the worst performance. The weakly-directional optimization of MaxDen rejects all electrons in our sample at \SI{10}{\kevee}. In Figure~\ref{fig:ang_res}, we saw that fluorine has poor directionality at \SI{10}{\kevee}, so it is no surprise that the weakly-directional optimization is superior throughout our energy range. For helium recoils, the weakly-directional optimization of MaxDen is better at lower energies below \SI{8}{\kevee}. At higher energies, the directional optimization wins. On the other hand, for ClustThres, we see that the performance is not sensitive to the optimization, for both fluorine and helium recoils.

Similarly to Section~\ref{Non-directional}, we compute a combined observable for each energy bin, except this time we only show results corresponding to a nuclear recoil selection efficiency of $\epsilon=0.5$. The observable optimizations that are included in our calculation of the combined observable depend on the energy bin and recoil species. For the fluorine recoils shown in Figure~\ref{f_rec}, the non-directional MaxDen optimization outperforms the directional optimization throughout the energy range, therefore only non-directional optimizations are used in the combined observable. On the other hand, for the helium recoils shown in Figure~\ref{he_rec}, the non-directional optimization of MaxDen only outperforms directional optimization below \SI{8}{\kevee}. Hence the non-directional optimizations are used below \SI{8}{\kevee} and the directional optimizations are used for higher energies. If at any point, a single observable outperforms the combined observable, the combined observable is re-defined to equal said observable. This situation only arises at low energies when the non-directional optimization of MaxDen dominates. We note that this combined result should be viewed as a lower limit on the achievable electron rejection factor. Our simple methodology can be significantly improved with a dedicated study on using machine learning to combine our observables. For the fluorine recoils in Figure~\ref{fig:all_E}, the combined observable rejects all (roughly one million) electron recoils in the \SI{9}{\kevee} energy bin. For the helium recoils, only one electron (out of roughly one million) is kept in the \SI{11}{\kevee} energy bin. For both of these cases, and all following energies, $R \geq 10^{6}$.

\subsection{Discussion}
\label{Discussion}

In Sections~\ref{Non-directional}--~\ref{other}, we demonstrated improved electron rejection (with respect to $dE/dx$) for detector-independent simulations by introducing new observables for recoil identification. With these observables, we expect to improve electron rejection for any directional recoil detector that has high spatial granularity compared to typical recoil length 
and is currently using $dE/dx$ to reject electron recoils. The extent to which the electron rejection is improved will depend on the length-to-diffusion ratio of the recoil tracks, the spatial granularity of the detector, and other detector-specific effects. Important aspects to consider when reviewing electron rejection results are the energy scales, drift length, the gas mixture and the total density. Below, we briefly review relevant experimental work that may benefit from and can be compared against our findings. To help compare the different gases used in the relevant works cited, Table~\ref{tab:2} summarizes several gas properties, including the quantity $1 / ( \rho \cdot \sigma_T )$ discussed in Section~\ref{choice}.

\begin{table}[tbp]
\centering
\begin{tabular}{|p{40mm}||p{13mm}|p{20mm}|p{17mm}|p{16mm}|p{15mm}|}
\hline
Gas Mixture & Pressure $\left[ \rm{Torr} \right]$ & Density $\left[ \rm{g} / \rm{cm}^3 \right]$ & Drift field $\left[ \rm{V} / \rm{cm} \right]$ & $\sigma_T$ $\left[ \si\micro \rm{m} / \sqrt{\textrm cm}  \right]$ & $1 / ( \rho \cdot \sigma_T )$  $\left[ \frac{ \rm{cm}^{7/2} }{  \rm{g} \cdot \si\micro \rm{m} } \right]$\\
\hline
 $80 \% \textrm{ He} + 10 \% \textrm{ CF}_4 + 10 \% \textrm{ CHF}_3$~[this work]& 60 & $6.69 \times 10^{-5} $ & 40.6  & 398  & 37.6  \\
 \hline
 $97.4 \% \textrm{ He} + 2.6 \% \textrm{SF}_6  $ \cite{cygnus} & 760 & $3.35 \times 10^{-4}$ & -- & 78.6   & 38.0 \\
 \hline
$\textrm{ CF}_4 $~\cite{dinesh} & 100  & $5.17\times 10^{-4}$ & 400 & 184 & 10.5 \\
\hline
 $ 70 \% \textrm{ CF}_4 + 28 \% \textrm{ CHF}_3  + 2 \% \textrm{ C}_4\textrm{H}_10$~\cite{MIMAC2} & 37.5  & $1.81\times 10^{-4}$ & 180 & 253 & 21.8 \\
\hline
 $ 60 \% \textrm{ He} + 40 \% \textrm{ CF}_4$~\cite{CYGNO_erej} & 760  & $1.68\times 10^{-3}$ & 500 & 150 & 3.97 \\
\hline
\end{tabular}
\caption{\label{tab:2} Properties of our gas mixture as well as the mixtures from refs.~\cite{cygnus,dinesh,MIMAC2,CYGNO_erej}. The transverse diffusion $(\sigma_T)$ was calculated via \texttt{Magboltz}~\cite{magboltz} for each gas. As discussed in Section~\ref{choice}, the $1 / ( \rho \cdot \sigma_T )$  value strongly impacts electron rejection capabilities.}
\end{table}

Using only projected length ($dE/dx$), N.S. Phan et al. rejected 25,696 Cobalt-60 induced electron recoils down to $\SI{10}{\kevee}$~\cite{dinesh}. Their detector used 100 Torr $\textrm{ CF}_4$ gas with a maximum drift length of \SI{10}{mm}. Although below \SI{10}{\kevee} they were no longer able to distinguish electrons recoils from nuclear recoils, their detection threshold is as low as 2 keVee suggesting their sensitivity can be improved with better electron rejection.

The MIMAC collaboration achieved an electron rejection of $R=10^5$ with an $86.49 \pm 0.17\%$  nuclear recoil efficiency on experimental fast neutron data~\cite{MIMAC2} (for BDT cut 0.189 in ref.~\cite{MIMAC2}). In this study, MIMAC combined a suite of observables, based on track topology and signal-pulse shape, together with a boosted decision tree. A monochromatic neutron field was used to create nuclear recoils (with fluorine recoil ionization energies up to 57 keV) and electron recoils with energies up to \SI{60}{\kevee} were created by introducing lithium-7 to their target. In our study, we present electron rejection for specific energy bins, whereas the MIMAC result is over their entire energy range. A plot of efficiency versus energy for different BDT cuts is presented in Figure 20 of ref.~\cite{MIMAC2}. Looking at BDT cut 0.189, the efficiency is below 0.1 up to an ionization energy of \SI{5}{keV}. The efficiency increases with energy and is approximately 1 beyond \SI{15}{keV}. Furthermore, their detector had a maximum drift length of \SI{18}{cm} and utilized a $70 \% \textrm{CF}_4 + 28 \% \textrm{ CHF}_3 + 2  \% \textrm{ C}_4\textrm{H}_{10}$  gas mixture at total pressure of 37.5 Torr.

The CYGNO collaboration recently conducted an experimental electron rejection study where the energy density was used to distinguish the electron recoils from nuclear recoils~\cite{CYGNO_erej}. CYGNO’s detector ran with a $60 \% \textrm{ He} + 40 \% \textrm{ CF}_4$  gas mixture at atmospheric pressure. The maximum drift length in the detector was \SI{20}{cm}. In this study, they identified \SI{5.9}{keV} electron recoils with $96.5\%$ efficiency at a nuclear recoil efficiency of $50\%$. With respect to our definition of electron rejection, this corresponds to $R = 28.6$.

Our simulation results suggest these experiments, which all utilize high-resolution readouts, and future experiments with similar capabilities, may be capable of significantly improved electron rejection at low energies. In particular, further advances should be possible with gas mixtures specifically optimized for electron rejection.

\section{Conclusion}
\label{sec:conclusion}

We have shown that if ionization in a TPC is detected with a highly-segmented charge readout (of order \SI{100}{\micro m}), physically intuitive variables based on the detailed charge cloud topology can be defined, significantly improving particle identification. These variables remain robust even below \SI{10}{\kevee} and are therefore of strong interest for dark matter detection. They are even robust in the most difficult regime for directional detectors, where recoils are short compared to diffusion, and directionality starts to fail. A simple attempt at combining these new variables into a single discriminant outperforms the traditionally used dE/dx by up to three orders of magnitude. Electron background rejection capabilities rise exponentially with energy: At 50\% nuclear recoil efficiency, more than 90\% of electrons, even at energies as low as \SI{3}{\kevee} -  \SI{4}{\kevee}, can be rejected. We achieve an electron rejection of $R \geq 10^6$ at energies equal to or greater than \SI{9}{\kevee} and \SI{11}{\kevee} for fluorine and helium recoils, respectively. 

Likely, a more sophisticated method of combining our observables could further improve electron rejection. We recommend a study of how electron rejection can be combined optimally via machine learning. Another near-term goal is the experimental verification of these predictions. Overall, these results are quite encouraging for the CYGNUS proposal to build a large nuclear recoil observatory, where the rejection of electron background at low energies has been identified as a critical issue.

\acknowledgments

MG would like to thank Stephen Biagi for useful discussions regarding \texttt{Degrad}. We would like to thank Jeffrey Schueler, Michael Hedges, and Brian Crow for their technical advice and Thomas Thorpe for his feedback on the manuscript. Finally, we would like to thank the anonymous referee for their constructive feedback. This work was supported by the U.S.~Department of Energy (DOE) via Award Number DE-SC0010504.


\bibliographystyle{JHEP} 
\bibliography{main}



\end{document}